\documentclass[twocolumn]{aastex61}

\usepackage{multirow}
\usepackage{isotope}
\newcommand{\eqt}[1]{Eq.\,(\ref{#1})}
\newcommand{\sect}[1]{Sect.\,\ref{#1}}
\newcommand{\tab}[1]{Table\,\ref{#1}}
\newcommand{\code}[1]{\texttt{#1}}
\newcommand{\mzams}{\ensuremath{M_{\rm ZAMS}}}

\newcommand{\ekinw}{\ensuremath{\, E_{\rm wind}}}
\newcommand{\teff}{\ensuremath{T_{\rm eff}}}

\newcommand{\msun}{\ensuremath{\, M_\odot}}

\submitjournal{ApJS}

\shorttitle{SYGMA}
\shortauthors{Ritter et al.}

\begin{document}

\title{SYGMA: Stellar Yields for Galactic Modeling Applications}

%%%%%%%%%%%%%
\correspondingauthor{Christian Ritter}
\email{critter@uvic.ca}

\author{Christian Ritter}
\affil{Department of Physics and Astronomy, University of Victoria, Victoria, BC, V8P5C2, Canada}
\affiliation{Keele University, Keele, Staffordshire ST5 5BG, United Kingdom}
\affiliation{JINA-CEE, Michigan State University, East Lansing, MI, 48823, USA}
\affiliation{NuGrid collaboration, \url{http://www.nugridstars.org}}

\author[0000-0002-9986-8816]{Benoit C\^ot\'e}
\affiliation{Konkoly Observatory, Research Centre for Astronomy and Earth Sciences, Hungarian Academy of Sciences, Konkoly Thege Miklos ut 15-17, H-1121 Budapest, Hungary}
\affiliation{National Superconducting Cyclotron Laboratory, Michigan State University, MI, 48823, USA}
\affiliation{JINA-CEE, Michigan State University, East Lansing, MI, 48823, USA}
\affiliation{NuGrid collaboration, \url{http://www.nugridstars.org}}

\author[0000-0001-8087-9278]{Falk Herwig}
\affiliation{Department of Physics and Astronomy, University of Victoria, Victoria, BC, V8P5C2, Canada}
\affiliation{JINA-CEE, Michigan State University, East Lansing, MI, 48823, USA}
\affiliation{NuGrid collaboration, \url{http://www.nugridstars.org}}

\author{Julio F. Navarro}
\affiliation{Department of Physics and Astronomy, University of Victoria, Victoria, BC, V8P5C2, Canada}

\author[0000-0003-2624-0056]{Chris L. Fryer}
\affiliation{Computational Physics and Methods (CCS-2), LANL, Los Alamos, NM, 87545, USA}
\affiliation{NuGrid collaboration, \url{http://www.nugridstars.org}}

%%%%%%%%%%%%%

\begin{abstract}
The Stellar Yields for Galactic Modeling Applications (\code{SYGMA})
  code is an open-source module that models the chemical ejecta and
  feedback of simple stellar populations (SSPs). It is intended for
  use in hydrodynamical simulations and semi-analytic models of
  galactic chemical evolution. The module includes the enrichment from
  asymptotic giant branch (AGB) stars, massive stars, Type-Ia supernovae (SNe~Ia) and
  compact binary mergers. An extensive and extendable stellar yields
  library includes the NuGrid yields with all elements and many
  isotopes up to Bi.  Stellar feedback from mechanic and
  frequency-dependent radiative luminosities are computed based on
  NuGrid stellar models and their synthetic spectra. The module
  further allows for customizable initial-mass functions and SN~Ia delay-time distributions to calculate time-dependent
  ejecta based on stellar yield input. A variety of r-process sites can be included. 
A comparison of SSP ejecta based on NuGrid yields with those from Portinari et al. (1998) and Marigo (2001) reveals up to a factor of 3.5 and 4.8 less C and N enrichment from AGB stars at low metallicity, a result we attribute to NuGrid's modeling of hot-bottom burning. Different core-collapse supernova explosion and fallback prescriptions
may lead to substantial variations for the accumulated ejecta of C, O
and Si in the first $10^7\, \mathrm{yr}$ at $Z=0.001$.  An online
interface of the open-source \code{SYGMA} module enables interactive
simulations, analysis and data extraction of the evolution of all
species formed by the evolution of simple stellar populations.
\end{abstract}


\def\aligned{\vcenter\bgroup\let\\\cr
\halign\bgroup&\hfil${}##{}$&${}##{}$\hfil\cr}
\def\endaligned{\crcr\egroup\egroup}

% !TEX root = ./paper.tex
\section{Introduction}\label{s.introduction}

\noindent Galactic chemical evolution (GCE) models \citep[see,
  e.g.,][]{audouze76,arnett96} require as input the time-dependent
nucleosynthetic output returned to the interstellar medium by evolving
stars and stellar explosions.  These metallicity-dependent yields are
combined to represent a simple stellar population (SSP), and
includes material processed by
low-mass stars, massive star winds, supernovae, and merger events.
SSPs are the core building blocks of galactic chemical evolution
models, and are used as input in hydrodynamic cosmological structure
and galaxy formation simulations
\citep[e.g.\ ][]{wiersma:09,2014MNRAS.444.3845F,2015ARA&A..53...51S}.

In this paper, we present the open-source, time-dependent SSP module
Stellar Yields for Galactic Modeling Applications (\code{SYGMA}). It
implements easily-modifiable prescriptions for the initial mass function (IMF), stellar
lifetimes, wind mass-loss histories and merger timescales.  It can also be
used in a stand-alone manner to assess the impact of stellar evolution
and nucleosynthesis modeling assumptions.  The concept of an
  SSP module is not new, and several alternative SSP codes can be
  found in the literature (e.g.,
  \citealt{1995PhDT........26G,1997MNRAS.290..471G,leitherer:99,kawata:03,few:12,2014MNRAS.444.3845F,2017A&A...605A..59R,2017AJ....153...85S}).

 \code{SYGMA} is part of the open-source python chemical evolution
  NuGrid framework
  \code{NuPyCEE}\footnote{\url{https://github.com/NuGrid/NuPyCEE}},
  includes examples and analysis tools, and represents the fundamental
  building-block component of our JINA-NuGrid chemical evolution
  pipeline (\citealt{cote:17c}).   This pipeline creates an integrated workflow that links basic stellar and
  nuclear physics investigations of the
  formation of elements in stars and stellar explosions, to SSPs, 
GCE and semi-analytic models
  (\citealt{cote:17}), and ultimately 
  hydrodynamic cosmological simulations.

\code{SYGMA} includes a number of published yield sets available as options in \code{NuPyCEE} which can be used to compare  \code{SYGMA} SSP models with those  using the latest set of NuGrid
yields \citep{pignatari:16,ritter:17}. The NuGrid yields have been derived from low-mass and massive star evolution and nucleosynthesis simulations created with the same codes for the entire mass range. All models adopt -- as much as possible  -- the same physics assumptions, including macro physics and nuclear reaction rates, for the entire mass range. This makes the NuGrid yield set the most internally consistent data set presently available.

The impact of yield uncertainties on GCE predictions has been shown in \cite{gibson:02} and \cite{romano:10}. 
\citet[][W09]{wiersma:09} developed a chemical feedback module for
hydrodynamic simulations, and found that their SSP ejecta can differ
by a factor of two or more for different yields available in
literature.  
%\code{SYGMA} has been used via NuGrid's \code{OMEGA} code
%\citep{cote:16,cote:17d,cote:17c,cote:17,cote:17b,ritter:17b}.

Aside from enriching the ISM, stars also alter the surrounding medium
through winds and radiation. These energy inputs into the ISM are the
basis of the stellar `feedback' introduced in most galaxy simulation
codes. A common approach is to adopt feedback prescriptions that
neglect their dependence on the stellar models from which the yields
are derived.  W09, for example, adopt a constant kinetic energy of
$10^{51}\, \mathrm{erg}$ for stars above the zero-age main-sequence
mass of $\mzams=6\msun$.  \code{SYGMA} implements these energy
sources based on the stellar models and supernova explosions from the
same NuGrid data set from which the yields were derived.

We present the functionality of our code in \sect{s.science1}.  In
\sect{s.science2} we analyze the ejecta of SSPs at solar metallicity
and low metallicity based on the new NuGrid yields
\citep{pignatari:16,ritter:17} compared to the combined yield set of
AGB yields of \citet[][M01]{marigo:01} and massive star yields of
\citet[][P98]{portinari:98}.  Also discussed are the effects of
core-collapse mass-cut prescriptions on yields from massive star
models. In \sect{s.conclusions} we summarize our results. An appendix
provides information on code verification and online access.

% !TEX root = ./paper.tex
\section{Methods}\label{s.science1} 

 \code{SYGMA} is part of  \code{NuPyCEE} which, together with the yield
  data used in this paper, is available on
  GitHub\footnote{\url{https://github.com/NuGrid/NuPyCEE}}. 
  For this work, we use \code{NuPyCEE} version 3.0, which
  can be recovered via Zenodo\footnote{\url{https://doi.org/10.5281/zenodo.1288697}} (\citealt{christian_ritter_2018_1288697}).

\subsection{Simple Stellar Population Mechanics}

A simple stellar population consists of an ensemble of stars of common age and metallicity.
 Chemical evolution assumptions describe
properties of the stellar population such as the number of stars
formed in a initial mass range.  Results shown in this section are
based on NuGrid yields
\citep{ritter:17}, Type~Ia supernovae (SNe~Ia) yields from
\cite{thielemann:86} and NS merger yields from \cite{rosswog:14}.

\subsubsection{Simple Stellar Population Ejecta}

\begin{figure*}[!htbp]
\centering
\includegraphics[width=1.\textwidth]{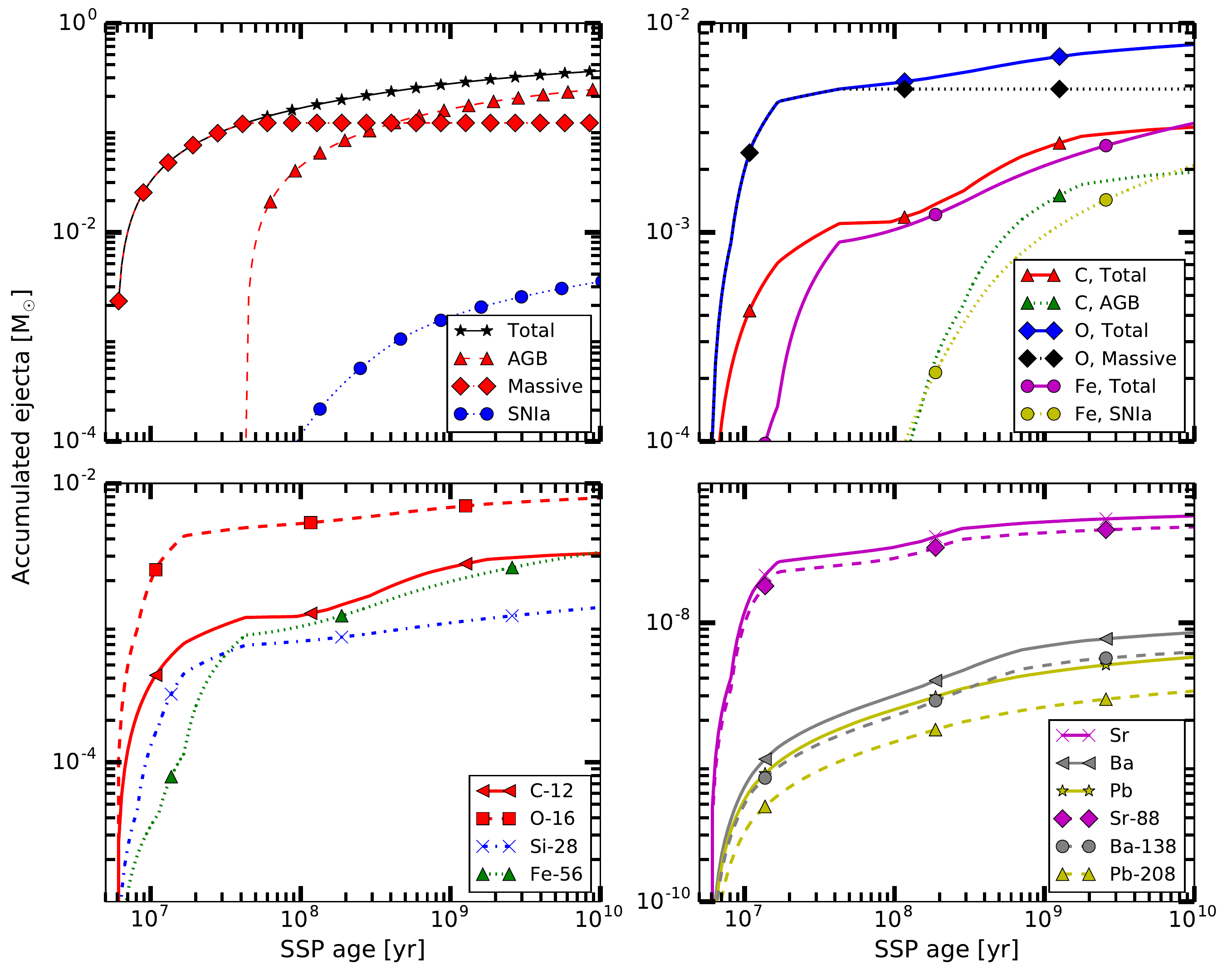}
\caption{Accumulated ejecta from AGB stars, massive stars and SNe~Ia
  for a SSP at $Z=0.02$ (top, left).  The mass of the SSP is
    set to 1\msun\ so that the SSP ejecta represents the mass ejected
    per units of stellar mass formed.  Accumulated ejecta of C, O and
  Fe from all (total) or from distinct sources (top, right).  Ejecta
  from of abundant isotopes of C, O, Si and Fe (bottom, left). Total
  accumulated ejecta of isotopes and elements of intermediate mass and
  from the first, second and third s-process peak (bottom, right).
  Access and interactive exploration of the figures is possible with
  the WENDI interface (Appendix\ \ref{s.appendixA}).}
\label{fig:chem_evol.png}
\end{figure*}

The cumulative contributions from AGB stars, massive stars and SNe~Ia
are tracked separately (Figure~\ref{fig:chem_evol.png}).  \code{SYGMA}
provides analytic tools to identify the most relevant nuclear
production site for a given element or isotope. For example, in an SSP
at $Z=0.02$, AGB stars (Figure~\ref{fig:chem_evol.png}) produce about
64\% of the total amount of C \citep[e.g.][]{schneider:14}, while
massive stars produce 62\% of the total amount of O \citep[e.g.][]{woosley:02}.  SNe~Ia produce 63\% of Fe
\citep[e.g.][]{thielemann:86}.  All stable elements and most isotopes
up to Bi are tracked (Figure~\ref{fig:chem_evol.png}).  To model the
element enrichment through other sources such as r-process from
neutrino-driven winds additional yields can be included as extra yield
tables.

\code{SYGMA} adopts the delayed production approximation
\citep{pagel:09}.  The stellar lifetimes are based on the same stellar
evolution models as the yields.  For AGB models they are the stellar
lifetimes until the end of the computation during the thermal-pulse
AGB stage or until the end of the post-AGB stage.  For massive star
models the time span is the time until core collapse.  The time span
or lifetime $\tau$ is interpolated using a log-log spline fit of the
tabulated lifetimes and initial masses.

The mass $\Delta \mathrm{M}_{\rm SSP}$ lost by a SSP over the time
interval [$t$,$t$+$\Delta t$] is
\begin{equation}
	\Delta \mathrm{M}_{\rm SSP} =  \int_{t}^{t+\Delta t} \xi(M_{\tau}(t^\prime)) \, M_{\star}(M_{\tau}(t^\prime),Z)\, dt^{\prime}
\end{equation} 
where $M_{\tau}(t)$ is the inverse of the lifetime function
$\tau(M)$. $\xi(M)$ is the IMF normalized to the
total stellar mass of the SSP and $M_{\star}(M,Z)$ is the total ejected
mass of the stars of mass $M$ and metallicity $Z$.

\subsubsection{Initial Mass Function}

The IMF $\xi$ gives the number of stars $N$
within an initial mass interval [$m_{1}$,$m_{2}$] via 
\begin{equation}
	 N = A \int_{m_{1}}^{m_{2}}{\xi(m^{\prime})}\, dm^{\prime}
\end{equation} 
where the normalization constant $A$ is derived from the total stellar
mass.  \code{SYGMA} provides a number of options \citep[see][for
  discussion]{kroupa:13}, such as the Salpeter IMF
(\citealt{salpeter:55})
\begin{equation}
	\xi(m) = A_S\, m^{-\alpha}
\end{equation} 
where $\alpha=2.35$ ($\alpha$ can be changed as input parameter), and
the Chabrier IMF (\citealt{chabrier:03})
\begin{equation}
\begin{aligned}
  	\xi(m) = &A_C\, m^{-1} e^{-\frac{(\log m - \log m_c)^2}{2 \sigma^2}}&\quad &\textrm{for}\ m \leq 1\msun, \\
	\xi(m) = &A_C\, m^{-2.3}&\quad &\textrm{for}\ m > 1\msun,
\end{aligned} 
\end{equation}
where $m_c=0.079$ and $\sigma=0.69$. The Kroupa IMF (\citealt{kroupa:01}) 
\begin{equation}
\begin{aligned}
	\xi(m) = &A_K\, m^{-0.3} &\quad &\textrm{for}\ 0.01\msun \leq m \leq 0.08\msun, \\
	\xi(m) = &A_K\, m^{-1.3} &\quad &\textrm{for}\ 0.08\msun \leq m \leq 0.50\msun, \\
	\xi(m) = &A_K\, m^{-2.3}  &\quad &\textrm{for}\  m \geq 0.50\msun.
\end{aligned} 
\end{equation}
is provided as well. A custom IMF can be defined as well in the code
and in the online version.  The lower and upper boundaries of the IMF
are input parameters, and related to the onset of H-burning and the
occurrence of the most massive stars respectively \citep[see
  discussion in][]{cote:16}.

\subsubsection{Type~Ia Supernova Rates}
\label{sn1arates}

The number of SNe Ia $N_{\rm Ia}$ per unit of $\msun$ formed in the time interval [$t$,$t$+$\Delta t$]
is given by
\begin{equation}
	 N_{\rm Ia} = A_{\rm Ia} \int_{t}^{t+\Delta t}
         f_{WD}(t^{\prime})\ \Psi_{\rm Ia}(t^{\prime})\, dt^{\prime}
\end{equation}
where $A_{\rm Ia}$ is a normalization constant, $f_{\rm WD}(t)$ is the
fraction of white dwarfs and $\Psi_{\rm Ia}(t)$ is the delay-time
distribution (DTD) at time $t$ (W09). White dwarfs with initial masses
between $3\msun$ and $8\msun$ are potential SN~Ia progenitors, as
commonly adopted \citep[e.g.][W09]{dahlen:04,mannucci:06}.  Either
$A_{\rm Ia}$ or the total number of SNe Ia per $\msun$ are required as
an input.  The power-law DTD of \cite{maoz:12}
\begin{equation}
	\Psi_{\rm Ia}(t) = t^{-1}
\end{equation}
can be selected, and 
the power-law exponent can be an input parameter.
An exponential DTD in the form of
\begin{equation}
	\Psi_{\rm Ia}(t) = \frac{e^{-t/\tau_{\rm Ia}}}{\tau_{\rm Ia}}
\end{equation}
and a Gaussian DTD as
\begin{equation}	
	\Psi_{\rm Ia}(t) =
        \frac{1}{\sqrt{2\pi\sigma^{2}}}e^{-\frac{(t-\tau_{\rm
              Ia})^{2}}{2\sigma^{2}}}
\end{equation}
are other options, as in W09. $\tau_{\rm Ia}$ is the characteristic
delay time and $\sigma$ defines the width of the Gaussian
distribution.  For a discussion of the parameters see W09 and
references within.  More generally, arbitrary delay-time distribution
functions may be used with the delayed-extra source option\footnote{\url{https://github.com/NuGrid/NuPyCEE/blob/master/DOC/Capabilities/Delayed_extra_sources.ipynb}} to explore new scenarios or to account for the individual contribution of different SN Ia channels.

\subsubsection{Neutron Star Merger Rates}

The number of NS mergers $N_{\rm NS}$ in the time interval
[$t$,$t$+$\Delta t$] is given as
\begin{equation}
	 N_{\rm NS} = A_{\rm NS} \int_{t}^{t+\Delta t} \Psi_{\rm
           NS}(t^{\prime},Z) \, dt^{\prime} \label{NSeq}
\end{equation}
where $A_{\rm NS}$ is a normalization constant and $\Psi_{\rm
  NS}(t,Z)$ is the NS DTD.  The user can choose between a power law
DTD, the DTD of \cite{dominik:12}, or a constant coalescence time.
For the DTD based on \cite{dominik:12} we adopt for the metallicity of
$Z\geq 0.019$ their DTD at solar metallicity and for $Z\leq 0.002$
their DTD at a tenth of solar metallicity. We interpolate the
distribution between these two metallicity boundaries. The
normalization constant $A_{\rm NS}$ is derived from the total number
of merger systems $N_{\rm NS,tot}$ via
\begin{equation}
	 A_{\rm NS}= \frac{N_{\rm NS,tot}}{ \int_0^\infty   \Psi_{\rm NS}(t,Z) \, dt^{\prime}}.
\end{equation}	 
$N_{\rm NS,tot}$ is calculated as  
\begin{equation}
	N_{\rm NS,tot} = 0.5\, f_m\, f_{b}   \int_{m_{1}}^{m_{2}}  \xi(m^{\prime}) \, dm^{\prime}
\end{equation}
where $f_m$ is the fraction of merger of massive-star binary systems
and $f_b$ is the binary fraction of all massive stars.  $f_m$ and
$f_b$ as well as the initial mass range for potential merger
progenitors [$m_1$,$m_2$] need to be provided as an input.  For
normalizing the number of NS merger with \eqt{NSeq}, the user can also
directly normalize the rate by providing the total number of NS merger
per unit of $\msun$ formed as an input.

With \code{SYGMA}, the DTD of NS mergers can also be defined
  by a simple power law.  In that case, the power-law index and the
  minimum and maximum coalescence timescales must be provided as input
  parameters.  Alternatively, arbitrary DTD functions can be assigned
  to NS mergers using our delayed-extra source
  implementation\footnote{\url{https://github.com/NuGrid/NuPyCEE/blob/master/DOC/Capabilities/Delayed_extra_sources.ipynb}}.
  This option can also be used for black hole neutron star mergers and
  for different SN~Ia channels.

\subsubsection{Yield Implementation} \label{netyieldeq}

The total yields $y_{\mathrm{tot},i}$ are defined as the total mass of element/isotope $i$ ejected over the lifetime of the star
and given as
\begin{equation}
y_{\mathrm{tot},i} = y_{0,i} + y_{n,i}
\end{equation}
where $y_{0,i}$ refers to the mass of element/isotope $i$ initially
available in the stellar simulation.  The net yields $y_{n,i}$ is the
produced or destroyed mass of element/isotope $i$. Yield tables with
total yields are the default input for \code{SYGMA}.

With the application of total yields the assumption holds that the
initial abundance of the underlying stellar model $y_{0,i}$ represents
the gas composition of the chemical evolution simulation
$y_{0,\mathrm{sim},i}$ at the time of star formation.  If material is
unprocessed throughout stellar evolution the total ejecta is
$y_{\mathrm{tot},i} = y_{0,i}$ and the error
$\epsilon_i=y_{0,\mathrm{sim},i} - y_{0,i}$ propagates into the total
ejecta. This could lead with $\epsilon_j>0$ to the artificial
production of isotope $j$ (for example in the case of r-process
species) if it is based solely on the solar-scaled initial abundance.
For element/isotope $k$ of secondary nucleosynthesis origin $y_{n,k}$
depends strongly on $y_{0,k}$ and on the error $|\epsilon_k|$.

Net yields can be applied in the code to take into account
$y_{0,\mathrm{sim},i}$ and \code{SYGMA} calculates the total ejecta
$y_{\mathrm{tot},i}$ as
\begin{equation}
y_{tot,i} = y_{0,sim,i} + y_{n,i}.
\end{equation}
With the destruction of element/isotope $h$ ($y_{n,h}<0$) and
$\epsilon_h<0$ more of element/isotope $h$ is destroyed than initially
available and the error is $|\epsilon_h|$.  In this case the code sets
$y_{\mathrm{tot},h}=0$.  For SN~Ia only total yields can be used.

%%TODO fix following sentence
An initial mass interval  can be specified
to take into account material locked away in
massive stars through black hole formation.  This IMF yield range goes
by default from $\mzams=1\msun$ to $\mzams=30\msun$.  The IMF yield
range and the IMF range can both be set separately for stars of $Z>0$
and Pop III stars to take into account different types of star
formation.

The yields of a specific tabulated stellar model are applied
  to all stars included in a certain interval of initial stellar
  masses. The lower and upper boundaries of this interval represent
  half the distance to the next available lower- and higher-mass
  tabulated stellar model, respectively.  For a given initial mass in
  this interval, the yields of the selected tabulated stellar model
  are re-normalized according to the fitted relation between the
  initial stellar mass and the total ejected mass.

\subsubsection{Stellar Feedback}

Mass-loss rates from stellar winds, stellar luminosities of
different energy bands, and kinetic energies of SN of SSPs can be
modeled with additional input tables.

\paragraph{Stellar winds}

\begin{figure}[!htbp]
\centering
\includegraphics[width=0.47\textwidth]{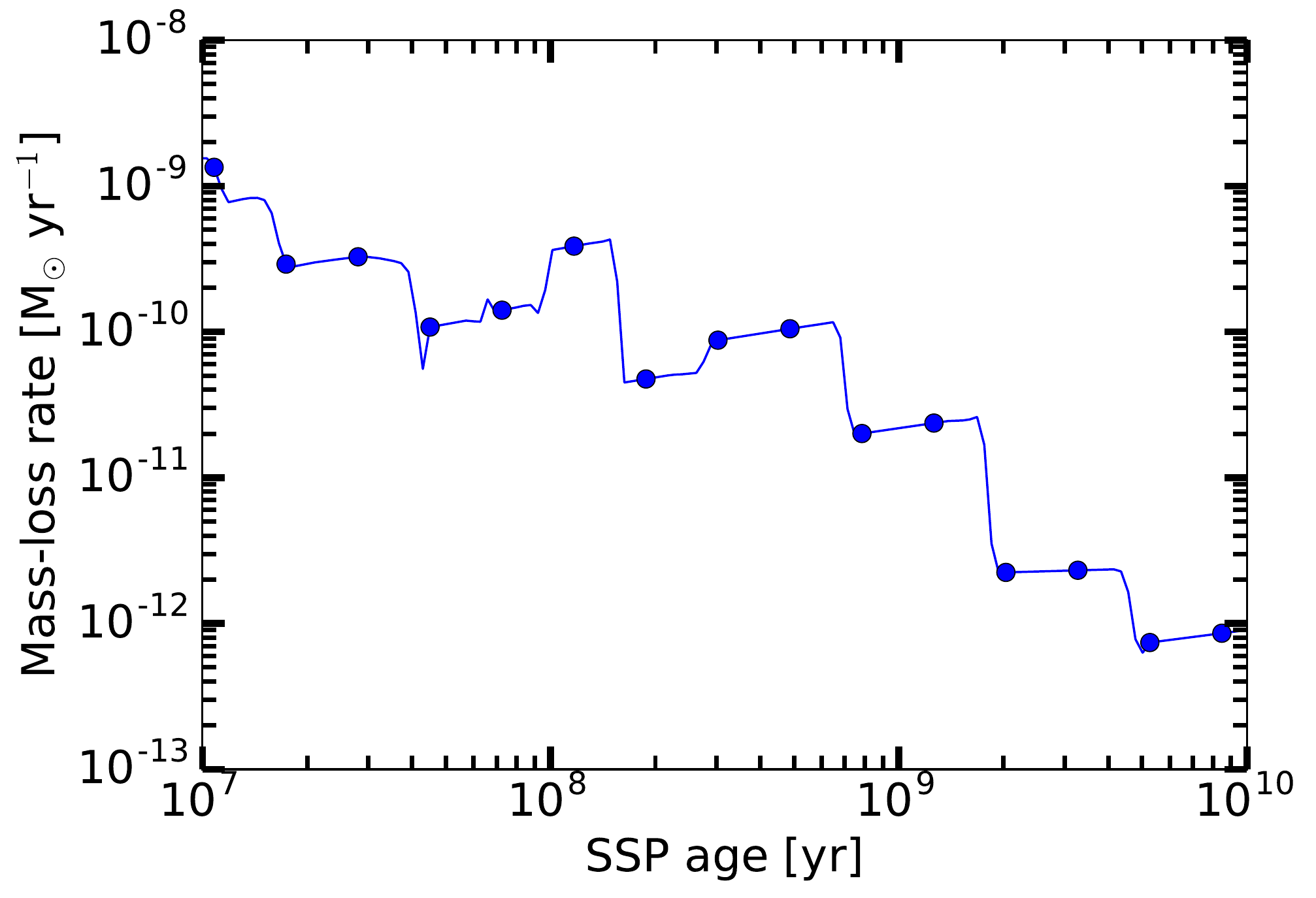}
\includegraphics[width=0.47\textwidth]{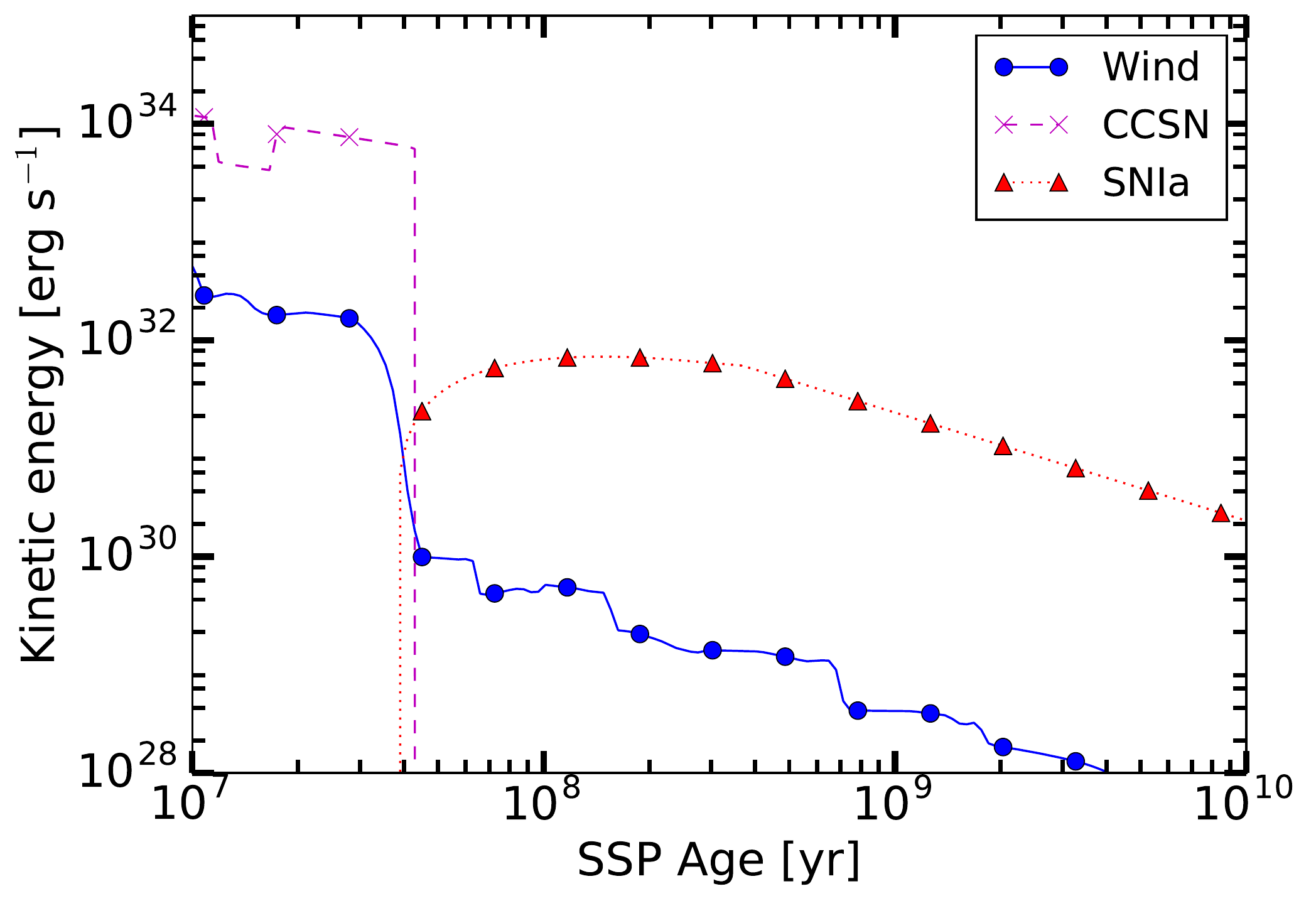}
\includegraphics[width=0.47\textwidth]{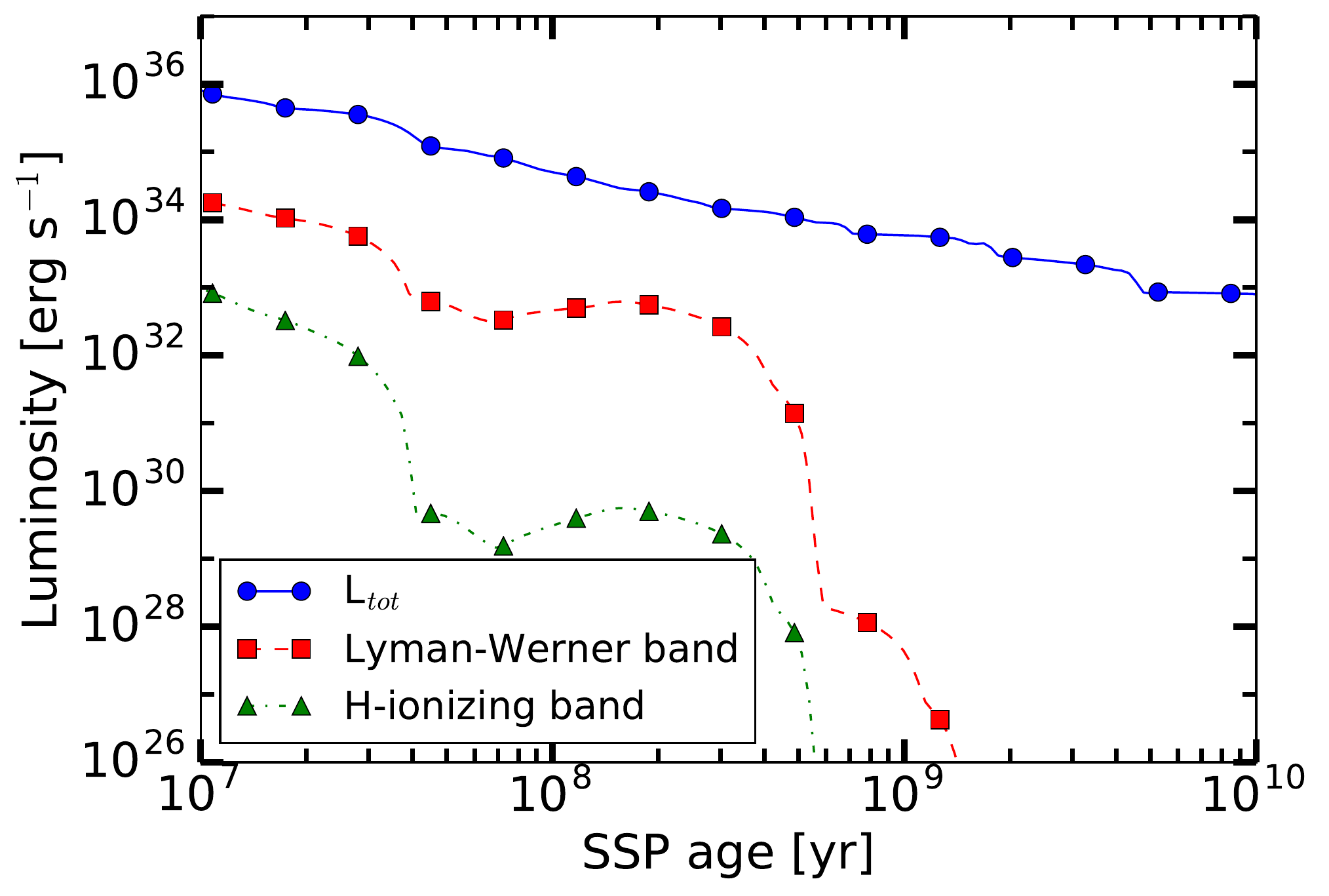}
\caption{Evolution of mass ejection of a SSP of $1\msun$ at $Z=0.02$ (top).  
Kinetic energies of stellar winds, CC~SNe and SNe~Ia (middle).
Time dependence of the total luminosity and luminosities in the Lyman-Werner and H-ionizing bands emitted by the SSP (bottom).
}
\label{fig:winds_parameter}
\end{figure}

The properties of stellar winds are of importance since these winds
can contribute to galactic winds which distribute metals and enrich
the intergalactic medium \citep{hopkins:12}.  We provide the time
evolution of the mass ejection rates of AGB and massive star models as
an input for \code{SYGMA}.  The mass-loss rates of SSPs
  decrease with time due to the declining AGB mass loss rates towards
  lower initial masses.  The mass-loss rates shown in
  Figure~\ref{fig:winds_parameter} for a SSP of $1\msun$ at solar
  metallicity are similar to Figures~3 and 4 in \cite{cote:15b}.
  Throughout this paper, we choose $1\msun$ for the mass of our SSPs
  so that the stellar ejecta represents the normalized mass ejected
  per units of stellar mass formed.  When comparing with
  \cite{cote:15b}, which adopted SSPs of $10^6\msun$, our results need
  to be scaled up by a factor of $10^6$.

The steps in the evolution of the ejected mass originate from the
transition between grid points of the stellar model grid.  The
strongest AGB mass loss originates from the $5\msun$ stellar model, as
visible in the peak shortly after $10^8\, \mathrm{yr}$.  To determine
the kinetic energy of stellar winds $\ekinw$ we calculate for each
stellar model the time-dependent escape velocity $v_{\rm esc}$. 

After deriving the terminal velocity $v_{\infty}$ from $v_{\rm esc}$
one can calculate the kinetic energy as
\begin{equation}
\ekinw(t) = \frac{1}{2} \dot{m}(t) v_{\infty}(t)^2.
\end{equation}
\cite{abbott:78} found the relation $v_{\infty}=3 v_{\rm esc}$ for observed O, B, A and Wolf-Rayet stars 
via radiation-driven wind theory.
We apply this relation because $\ekinw$ originates mainly from the most massive stars with radiative winds \citep{leitherer:92}.

The stellar grid includes stellar models up to $\mzams=25\msun$, while stellar models at higher initial mass are expected to contribute 
the majority of the kinetic wind energy of the stellar population. 
Hence our kinetic energy of winds are similar to those of the SSP with upper IMF limit of $30\msun$ of \code{starburst99} \citep[Fig. 111,][]{leitherer:99}.

For AGB models we approximate $v_{\infty}$ as $v_{\infty}=v_{\rm esc}$ which might overestimate the kinetic energy contribution of the AGB phase as indicated by observations \citep{bolton:00}. Energetic contributions of AGB stars are often not considered in SSP models since their contribution is negligible compared to massive stars \citep[][]{leitherer:99,cote:15b}. Our largest kinetic energies of winds 
originate from massive star models while the kinetic energies of winds from AGB models are considerably smaller (Figure~\ref{fig:winds_parameter}).

%%%%%%
\begin{deluxetable*}{llllllll}
\tablewidth{0pc}
\tablecaption{Properties of yield sets of AGB and massive stars available in \code{SYGMA}. Shown are the NuGrid sets \citep{pignatari:16,ritter:17} as well as low mass /massive star yield combinations M01P98 \citep{marigo:01b,portinari:98}, K10K06 \citep{kobayashi:06,karakas:10} and C15K06 \citep{cristallo:15,kobayashi:06}.
Properties are the initial mass and metallicity range, available S-AGB models, the modeled heavy-element processes,
the heaviest available stable element in the network provided and the modeling of the CCSN explosion. 
\label{tab:yieldcompilations}}
\tablehead{ \colhead{ID} & \colhead{Masses} & \colhead{Metallicity} & \colhead{S-AGB} & \colhead{Heavy-element processes} & \colhead{Network} & \colhead{Exp}}
\startdata
NuGrid & $1 - 25$   & $0.0001 - 0.02$   & yes & main s, weak s, p, $\gamma$ & Pb & yes \\
M01P98 & $1 - 120$ & $0.0004 - 0.05$   & no & - &Fe & no    \\
K10K06 & $1 - 40$   & $0.0001 - 0.02$ & no & - & Ni & yes  \\
C15K06 & $1.3 - 40$   & $0.0001 - 0.02$ & no & main s &Ge & yes  \\
\enddata
\end{deluxetable*}
%%%%%% 

\paragraph{Supernova energies}

The kinetic energies of core-collapse supernovae (CCSNe) is usually taken as $10^{51}\, \mathrm{erg}$, which is similar to the observed explosion energies of
SN such as SN 1987A \citep{arnett:89}. We apply CCSN energies based on the CCSN explosion prescription of \cite{fryer:12} used for NuGrid stellar models.
For each initial mass and metallicity of the stellar models the kinetic energy is extracted from Fig.\,2 of \cite{fryer:12}.
Energies for stellar models above $25\msun$ are not taken into account.  Such stellar models are not included in the NuGrid model grid because
they do not explode according to our remnant mass model.
The kinetic energy of a SNe Ia is an input parameter of \code{SYGMA} and is set to $10^{51}\, \mathrm{erg}$.

For a SSP of $1\msun$ at solar metallicity, the kinetic energy from CC~SN explosions of $\approx10^{34}\, \mathrm{erg\,s}^{-1}$
is similar to \code{starburst99} \citep[Fig.\ 113,][]{leitherer:99}. The kinetic energies from SNe~Ia  is more than $1\, \mathrm{dex}$
lower than the kinetic energy from CC~SN explosions (Figure~\ref{fig:winds_parameter}) and very similar to \citet[][Fig.\ 4]{cote:15b}.  
In \cite{leitherer:99} and \cite{cote:15b}, the mass of the SSPs are 10$^6\msun$, so our results need to be scaled up by a factor of 10$^6$
when doing the comparison.

\paragraph{Stellar luminosities}

Stellar radiation alters the surrounding medium through ionization and
radiation pressure.  \code{SYGMA} computes luminosities of SSPs based
on time-dependent bolometric luminosities of stellar models. The
latter are provided as table input.  The luminosities in specific
wavelength bands such as the H-ionizing band ($13.6\, \mathrm{eV}$ -
$24.6\, \mathrm{eV}$) are calculated as well as the time-dependent
luminosities of luminosity bands based on spectra of the stellar
evolution models. The latter are from the PHOENIX \citep{husser:13}
and ATLAS9 \citep{castelli:04} synthetic spectra libraries.

Spectra are derived from the best match of effective temperature
$\teff$, gravity $g$, [Fe/H] and $\alpha$-enhancement of a stellar
model. We adopt the $\alpha$-enhancement of the initial abundance of
the stellar models and neglect any changes of the surface abundance
during stellar evolution.  The $\alpha$-enhancement at low
metallicity of NuGrid models is based on observations of individual
elements and each element has its own enhancement. The corresponding
average $\alpha$-enhancement is [$\alpha$/Fe] = 0.8 by taking into
account the initial abundance of each element. %We The bolometric
The total luminosity, as well as the luminosity in two wavelength bands, of an SSP
of $1\msun$ at $Z=0.02$ are shown in the lower panel of
Figure~\ref{fig:winds_parameter}.

\subsection{Stellar Yields}

Metallicities for which stellar yields are provided can be selected as
initial metallicities of the SSP.  The number of isotopes
  included in SYGMA is flexible and is automatically set when reading
  the input yield tables.  The lifetime and final mass for each star
are required in the yield input tables.  In the following we introduce
the yield sets of our comprehensive yield library for
\code{SYGMA}. Additional yields can be added on request.

The default yields of AGB models, massive star models and
core-collapse supernova models are from the NuGrid collaboration and
include the metallicities $Z=0.02$, $0.01$, $0.006$, $0.001$, and
$0.0001$ \citep[yield sets NuGrid$_{\rm d/r/m}$,
  \tab{tab:yieldcompilations},][]{ritter:17}. Yields for twelve
stellar models between $\mzams=1\msun$ and $\mzams=25\msun$ are
provided for each metallicity, including super-AGB models.

To briefly summarize, the thermal pulse AGB (TP-AGB) stages of low-
and intermediate-mass stars and all burning stages of massive stars
until collapse are included.  All stable elements and many isotopes up
to Bi are provided in the yield tables.  Convective boundary mixing is
applied at all convection zones in AGB models.  A nested-network
approach resolves hot-bottom burning during post-processing
and predicts isotopes of the CNO cycle and s process isotopes.  As
part of the semi-analytic explosion prescription a mass- and
metallicity dependent fallback is applied which accounts for the
observed NS and BH mass distribution. Fallback in stellar models of
high initial mass leads to black hole formation.  Yield sets are
available for delayed explosions (NuGrid$_{\rm d}$), rapid explosions
(NuGrid$_{\rm r}$) and a combination of both (NuGrid$_{\rm m}$, see
\sect{s.science5}).  The NuGrid data is available online as tables and
through the WENDI interface (Appendix\ \ref{s.appendixA}).

The yield set of the Padua group (M01P98) consists of AGB yields from
M01 and massive star yields from P98.  Stellar yields for initial
masses between $\mzams=1\msun$ and $\mzams=120\msun$ and $Z=0.02,
0.008, 0.004, 0.0004$ are available.  The AGB stage of stellar models
up to $\mzams=5\msun$ is based on synthetic models which are
calibrated against observations. For stellar mass models with
$\mzams=6\msun$ and $\mzams=7\msun$ the AGB stage is not modeled.  The
contribution from CCSN nucleosynthesis of the massive star models is
derived from the explosions of massive star models of
\cite{woosley:95}.

AGB yields from \cite{karakas:10} and massive star yields from
\cite{kobayashi:06} are provided in the yield set K10K06.  Yields with
initial masses between $\mzams=1\msun$ and $\mzams=40\msun$ and for
$Z=0.02$, 0.008, 0.004, 0.001, and 0.0001 are included.  AGB yields
are available up to $\mzams=6\msun$. Stellar yields include elements
up to Ni and Ge respectively and we adopt elements up to Ni in the
K10K06 yield set.  Explosive nucleosynthesis is based on CCSN and
hypernova models with a mass cut set to limit the amount of ejected Fe
to $0.07\msun$.  Hypernova models adopt a mixing-fallback mechanism.

Yields of light and heavy elements for AGB models up to
$\mzams=6\msun$ are provided in the F.R.U.I.T.Y. database
\citep[e.g.][]{cristallo:15}.  We combine F.R.U.I.T.Y. yields with
those of K06 in the C15K06 yield set which includes $Z=0.02$, 0.008,
0.004, 0.001, and 0.0001.  Since yields of K06 are available only
until Ge the C15K06 yield set does not include the heavier elements
provided in the F.R.U.I.T.Y. database.

Additional yield sets can be created by combining other
yield tables.  Available options include CCSN yields, hypernova yields
and pair-instability supernova yields of \cite{nomoto:13}.
Zero-metallicity massive star yields are from \cite{heger:10}.  Yields
of magneto-hydrodynamic explosions of massive star models are from
\cite{nishimura:15}.  CCSN neutrino-driven wind yields are based on
simple trajectories of analytic models (N. Nishimura, private
communication).  r-process yields which are calculated with the
solar-system residual method are from \cite{arnould:07}.  Yield tables
of the dynamic ejecta of NS models which reproduce the strong
r-process component are from \cite{rosswog:14}.

Yields of 1D SN~Ia deflagration models are from \cite{thielemann:86},
\cite{iwamoto:99} and \cite{thielemann:03}.  Yields of
\cite{seitenzahl:13} are based on tracer particles of 3D hydrodynamic
simulations of delayed-detonation SNI~a models and are available for
progenitor metallicities of $Z=0.02$, 0.01, 0.002, and 0.0002.

% !TEX root = ./paper.tex

\section{Results}\label{s.science2}

%%%%%%%
\begin{deluxetable}{ll}
%\tablewidth{10.\textwidth}
\tablecaption{SSP parameters IMF range for \sect{s.science2} ($^*$)
  and Appendix~\ref{s.appendixveri} ($^+$), IMF type, characteristic delay time
  $\tau$, the normalization of the exponential delay-time
  distribution function of SNe~Ia $A_{\rm Ia}$, and the transition
  mass between AGB and massive stars $M_{\rm mass}$.
\label{tab:ce_parameter}}
\tablehead{ \colhead{Parameter} & \colhead{Adopted choice} 
}
\startdata
IMF type & Chabrier \\
IMF range & $0.1$ - $100\msun$ \\
IMF yield range$^*$ & $1$ - $30\msun$ \\
IMF yield range$^+$ & $0.8$ - $100\msun$ \\
$\tau_{\rm Ia}$ & $2\times10^{9}\, \mathrm{yr}$\\
$A_{\rm Ia}$ & 0.02 \\
$M_{\rm mass}$ & $8\msun$ \\
\enddata
\end{deluxetable}
%%%%%%%

\subsection{Simple Stellar Populations at Solar Metallicity}

\begin{figure*}[!htbp]
\centering
\includegraphics[width=1.\textwidth]{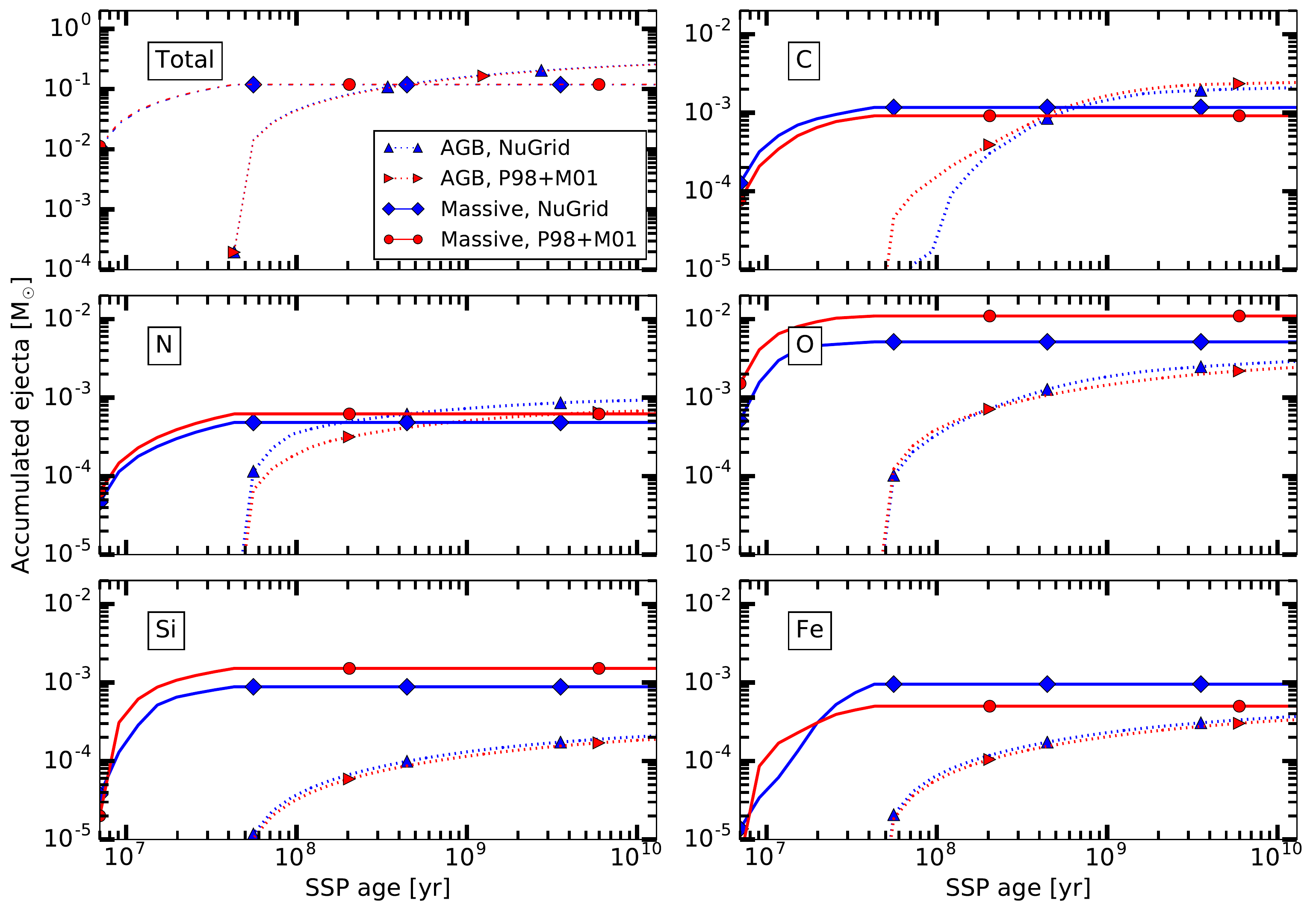}
\caption{Accumulated ejecta of AGB stars and massive stars based on NuGrid yields and M01P98 yields, for an SSP of $1\msun$ at $Z=0.02$.
}
\label{fig:sspsolarz}
\end{figure*}

We compare the ejecta of two SSPs based on the yield set NuGrid$_{\rm
  d}$ %\citep[delayed explosion prescription by][]{fryer:12} and the
yield set M01P98.  Applied are NuGrid yields at $Z=0.02$, P98 yields
at $Z=0.02$ and M01 yields at $Z=0.019$ which we will refer to as
yields at solar metallicity.  The initial abundances  of the
NuGrid yield set are selected.  We
calculate the total yields $y_{\rm tot}$ from the net yield set M01P98
(Sec.\ \ref{netyieldeq}).  We modified the M01P98 massive star yields by the
factors 0.5, 2 and 0.5 for C, Mg, and Fe, respectively, as done in
W09.  Those modifications are justified in Appendix A3.2 of
  W09 and are applied in our work to provide a consistent comparison
  with the results of W09.
%%TODO only for this particular case, right?!

In this section we use identical SSP parameters
(\tab{tab:ce_parameter}) to identify the differences due to different
yield sets from M01P98 and NuGrid.  $M_{\rm mass}$ represents
  the initial stellar mass that marks the transition between AGB and
  massive stars.  We choose $M_{\rm mass}$ in agreement with the
upper limit of the progenitors of SNe~Ia applied in the SSP code
(\sect{sn1arates}).  The evolution of the total accumulated mass of
AGB stars and massive stars is about the same for NuGrid yields and
M01P98 yields due similar amounts of total ejecta of both yield sets.

%This leads to the larger values for AGB stars in the SSP ejecta.
\begin{figure*}[!htbp]
\centering
\includegraphics[width=0.49\textwidth]{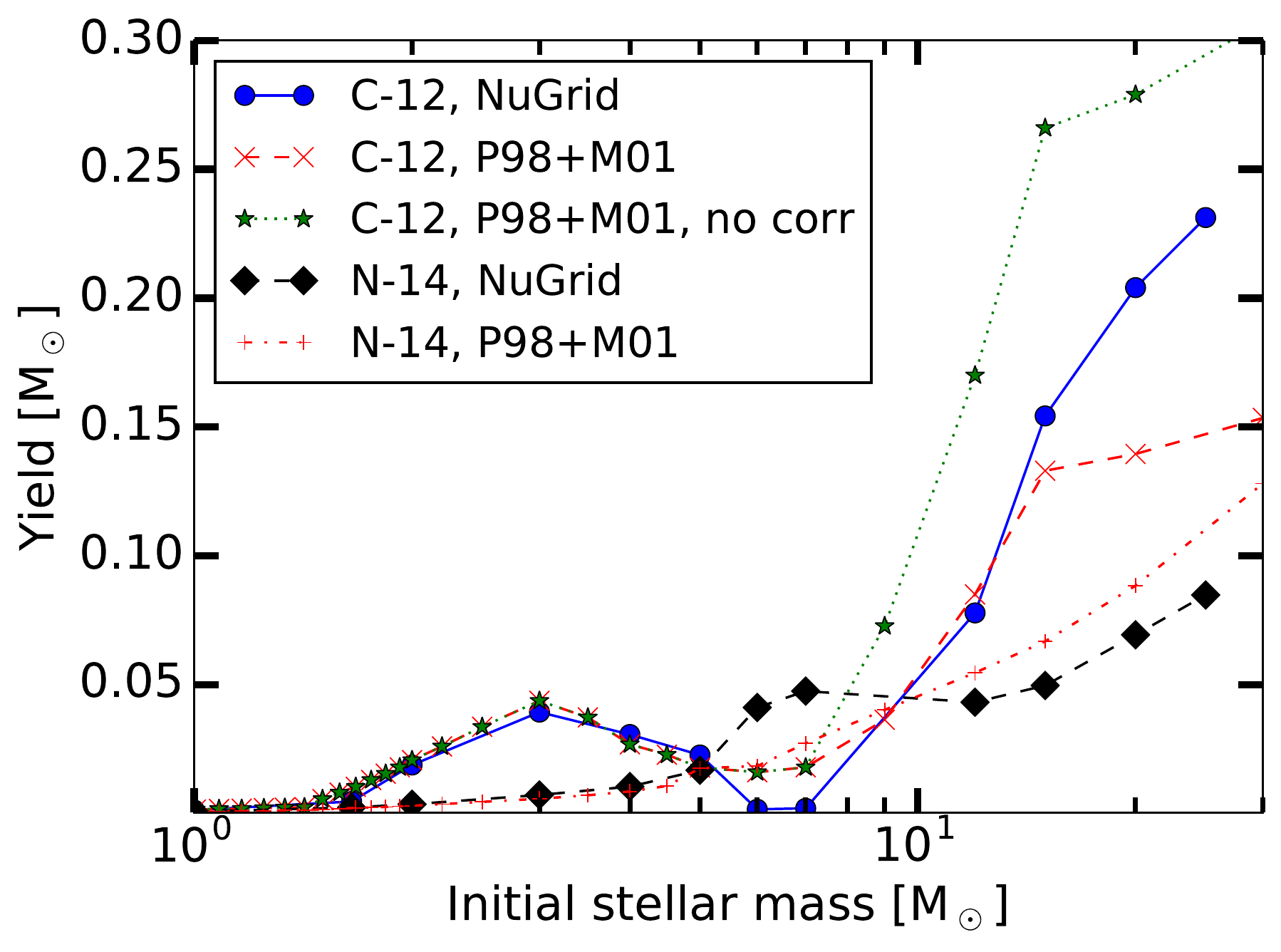} \hspace*{0.1cm}
\includegraphics[width=0.49\textwidth]{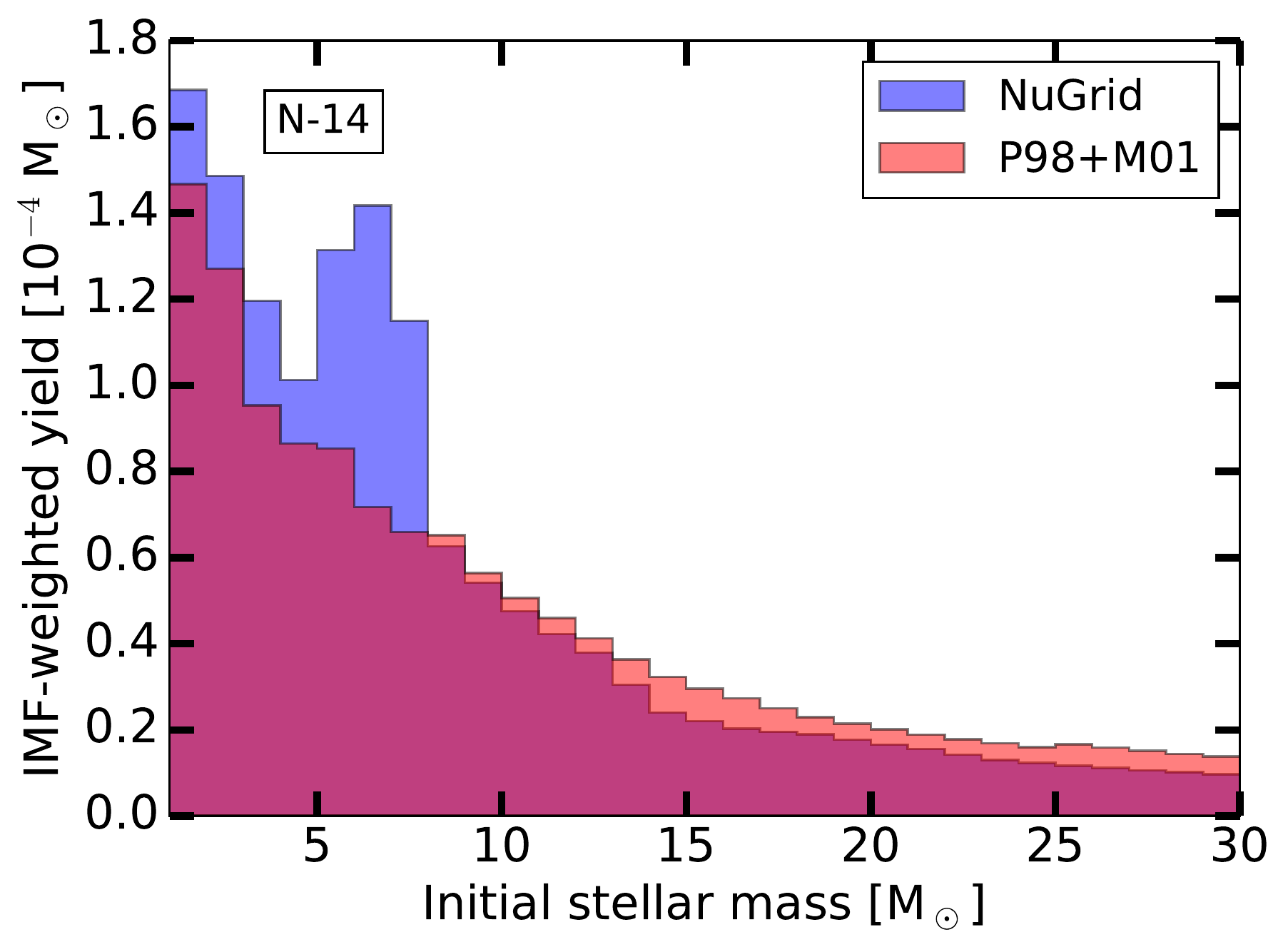}
\caption{\isotope[12]{C} and \isotope[14]{N} yields at solar metallicity versus initial mass of NuGrid, M01P98 and M01P98 without correction factor (no corr) of 0.5 (left). IMF-weighted yields of \isotope[14]{N} of stars of different initial mass (right).
%\href{https://github.com/NuGrid/NuPyCEE/blob/master/DOC/Papers/SYGMA_paper/Section3.ipynb}{Online Access}.
}
\label{fig:c_n_ce}
\end{figure*}

\paragraph{C, N} AGB stars are important sites of dust production \citep[e.g.][]{ferrarotti:06,schneider:14} and feature the primary production
of C \citep{herwig:05}. In massive and super-AGB stars C is transformed into N in 
hot-bottom burning (HBB) in the AGB stage \citep{lattanzio:96}. The TP-AGB phase changes dramatically the structure and chemistry of intermediate mass models.
Lower C yields in the NuGrid stellar models of $\mzams=6\msun$ and $\mzams=7\msun$ compared to stellar models of lower initial mass is due to more efficient \isotope[12]{C} destruction in the CNO cycle during HBB.
The stellar models with $\mzams=6\msun$ and $\mzams=7\msun$ by P98 do not include the TP-AGB phase, the third dredge-up (TDUP) and destruction of C during HBB.
The accumulated ejecta of \isotope[12]{C} of the SSP with M01P98 yields is higher than with NuGrid yields before $\approx2\times10^8\, \mathrm{yr}$ (Figure~\ref{fig:sspsolarz}).

While the total SSP ejecta of C from AGB stars are 11\% lower for NuGrid yields than for M01P98 yields, for massive stars total SSP ejecta are 30\% larger. The massive stars of the SSP with NuGrid yields produce at all times more C than those of the SSP with  M01P98 yields. Without the decrease of massive star yields of P98 by 0.5 as in this work and in W09 M01P98 yields would produce the most C.
In the latter case AGB stars would start to dominate the total production of C at $9\times10^{8}\, \mathrm{yr}$ instead of $3.6\times10^{8}\, \mathrm{yr}$.

HBB transforms \isotope[12]{C} into \isotope[14]{N}, leading to larger stellar yields of N from NuGrid compared to the models of M01P98 (Figure~\ref{fig:c_n_ce}). The
result is a bump in the IMF-weighted ejecta of stellar models with $\mzams=6\msun$ and $\mzams=7\msun$.
AGB stars based on NuGrid yields eject in total 28\% more N than those with M01P98 yields. For massive stars it is 18\% less N with NuGrid yields compared to M01P98 yields. 

\begin{figure*}
\centering
\includegraphics[width=0.465\textwidth]{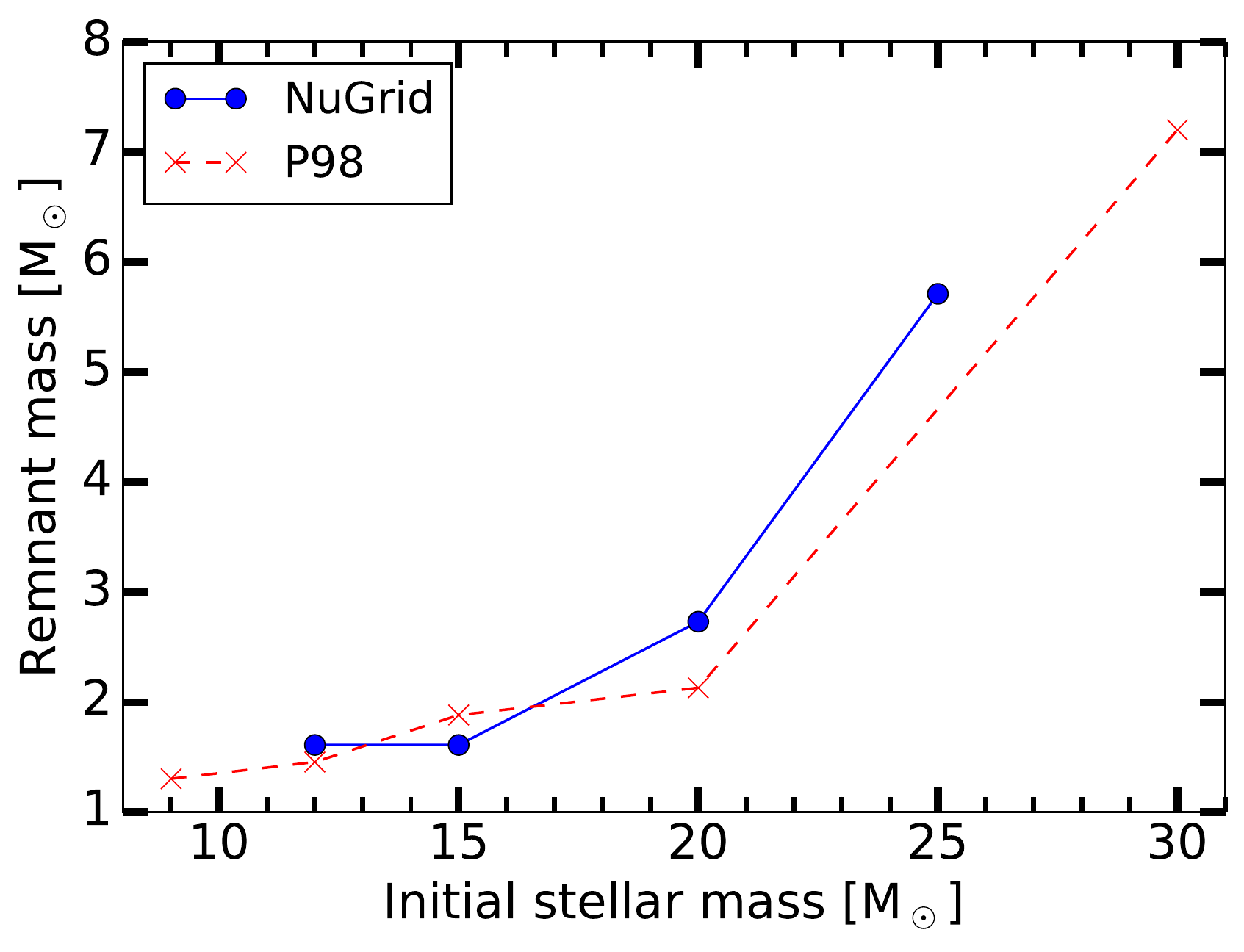}\hspace*{0.1cm}
\includegraphics[width=0.49\textwidth]{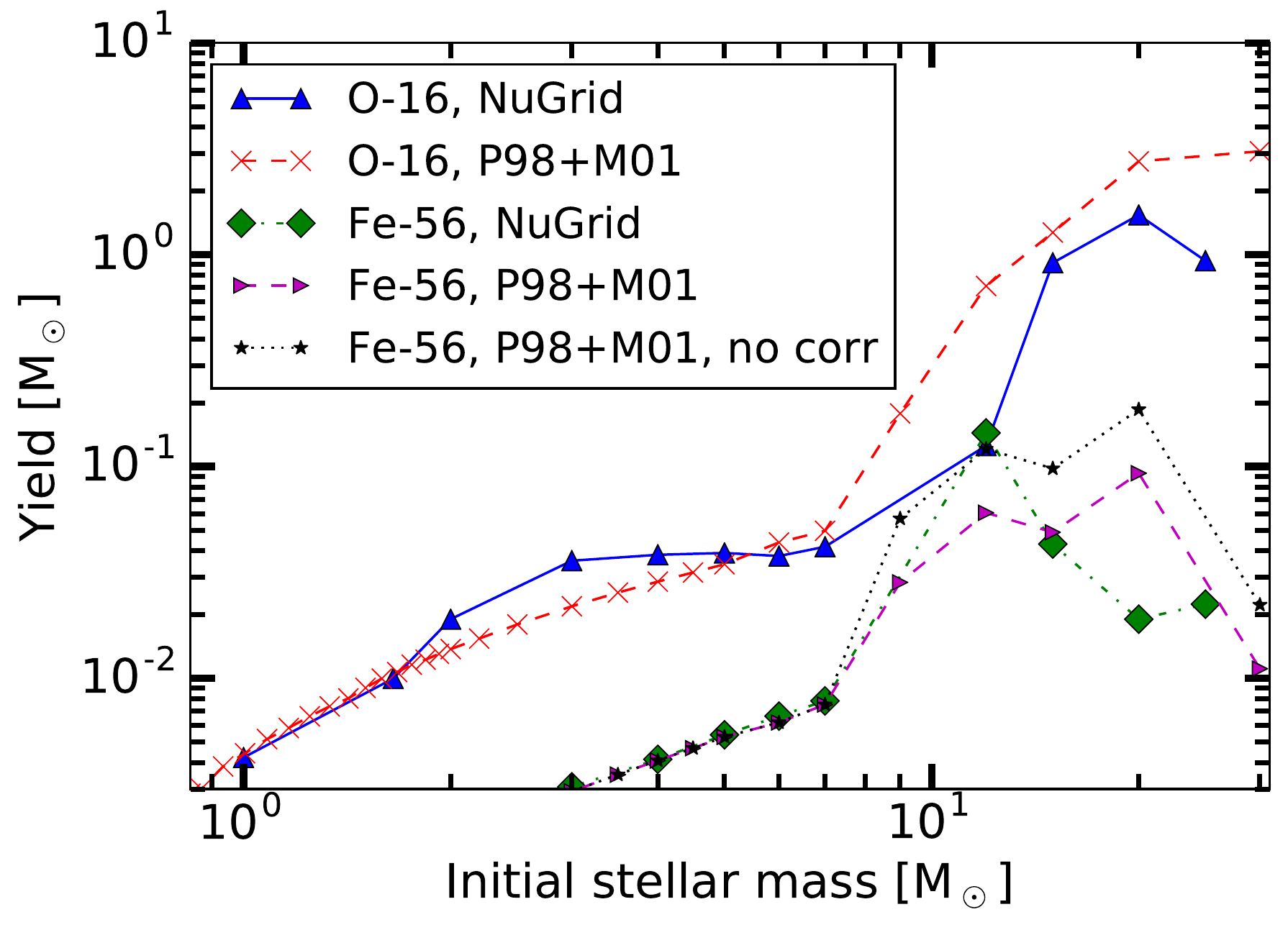}
\caption{Remnant masses versus initial mass of NuGrid and P98 models at solar metallicity (left).  Yields of O represented by \isotope[16]{O} and Fe yields represented by \isotope[56]{Fe} at solar metallicity versus initial mass (right).
}
\label{fig:o_ce}
\end{figure*}

\paragraph{O, Si} Most O is produced in massive stars \citep{timmes:95,woosley:02}. 
Larger O shells for more massive stars do not necessarily lead to larger amounts of O ejected when  strong fallback is taken into account.
In stellar models with $\mzams=20\msun$ and $\mzams=25\msun$ from NuGrid strong fallback of large parts of the O shell results in
larger remnant masses and less O ejection at high initial mass compared to P98 (Figure~\ref{fig:o_ce}). This results in 47\% lower O ejecta of the SSP than compared to the SSP with M01P98 yields (Figures~\ref{fig:o_ce} and \ref{fig:sspsolarz}).

AGB models of NuGrid include convective boundary mixing, which results in He-intershell enrichment of O of about 15\% in low-mass TP-AGB models compared to 2\% without any convective boundary mixing \citep{herwig:05}. 
M01 do not include any convective boundary mixing which is why the mass of O from AGB stars is higher with NuGrid yields than with M01 yields.
We find that the effect of convective boundary mixing on the total AGB production of O is small: the difference in SSP ejecta between NuGrid yields and M01 yields  is only 17\% (Figure~\ref{fig:sspsolarz}).

\begin{figure*}
\centering
\includegraphics[width=1.\textwidth]{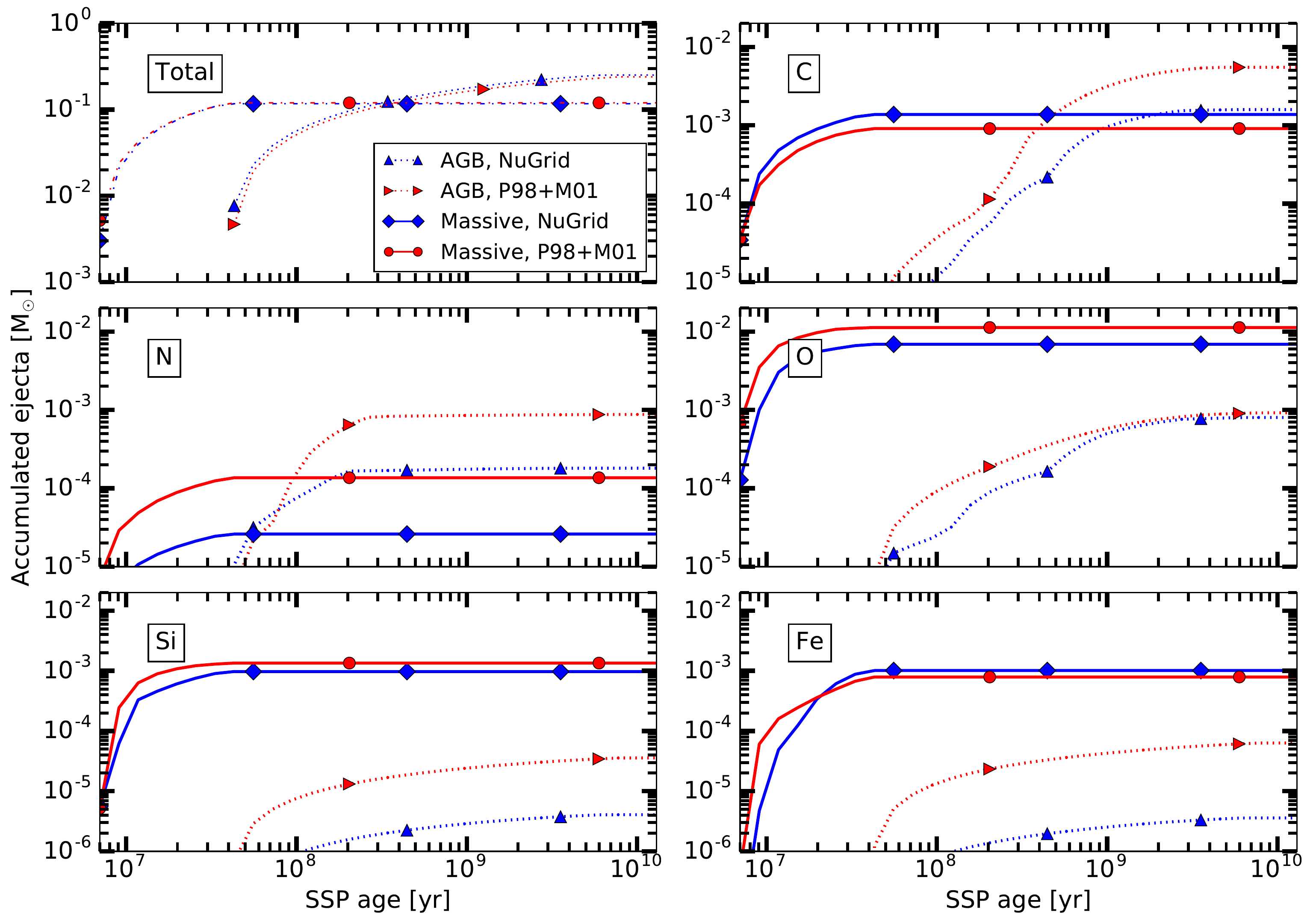}
\caption{Accumulated ejecta of AGB stars and massive stars based on NuGrid yields at $Z=0.001$ and M01P98 yields at $Z=0.004$, for an SSP of $1\msun$. 
}
\label{fig:ssplowz}
\end{figure*}

Si is produced in massive stars \citep[e.g.][]{timmes:95}.  Due to its
closer proximity to the core than lighter elements the difference in
chemical evolution of Si according to NuGrid and M01P98 yields are
only slightly smaller than for O. 43\% less Si is ejected
with NuGrid yields than with yields of M01P98.  AGB stars eject mostly
unprocessed Si and the total amount ejected is within 10\% between the
yield sets.

\paragraph{Fe}  %SNe Ia produce most of the Fe (e.g. \citealt{thielemann:86}) and are not discussed here. 
The SSP ejecta of Fe of massive stars are qualitatively different between the yield sets
(Figure~\ref{fig:sspsolarz}) due to the variation of Fe yields with initial mass (Figure~\ref{fig:o_ce}). This is primarily a consequence of the NuGrid remnant mass and fallback model adopted in this version of the NuGrid yields \citep{ritter:17}. As a result Fe ejecta are two times larger for the $12\msun$ model compared to the $15\msun$ model at $Z=0.02$ and the $20$ and $25\msun$ models have further reduced Fe yields. 
 M01P98 Fe yields peak, instead, at $\mzams=20\msun$. In cases like Fe, where the NuGrid yields predict a substantial rise toward the transition mass to white dwarf formation, the choice of interpolation vs.\ extrapolation of yields can be quite relevant. Our approach is described in  \sect{netyieldeq}. Ultimately, a higher model density at low masses would be the best solution. However, the lowest mass massive star models are the most difficult to calculate and are still subject to many modeling uncertainties.
At later times the SSP ejecta of Fe from massive stars based on NuGrid yields are larger and the total ejecta based on NuGrid is 47\% larger than the one based on M01P98 yields.
SSP ejecta of Fe of AGB stars is unprocessed and its total amount ejected is within 10\% between yield sets. 

\subsection{Simple Stellar Populations at Low Metallicity}

To analyze the SSP ejecta at low metallicity we compare the yield set
NuGrid$_{\rm d}$ at $Z=0.001$ with the yield set M01P98 at
$Z=0.004$. To calculate the total yields from M01P98 (see
\sect{netyieldeq}), we used the initial abundance adopted for
NuGrid models at solar Z but scaled down to $Z=0.004$.  Yields of both
sets at the same sub-solar metallicity are not available.  The initial
abundance of NuGrid yields at $Z=0.001$ is $\alpha$-enhanced in
contrast to P98 and M01.  We compare the total accumulated ejecta of
AGB stars and massive stars based on yields from M01P98 with those
based on NuGrid yields (Figure~\ref{fig:ssplowz}).

\begin{figure*}
\centering
\includegraphics[width=0.485\textwidth]{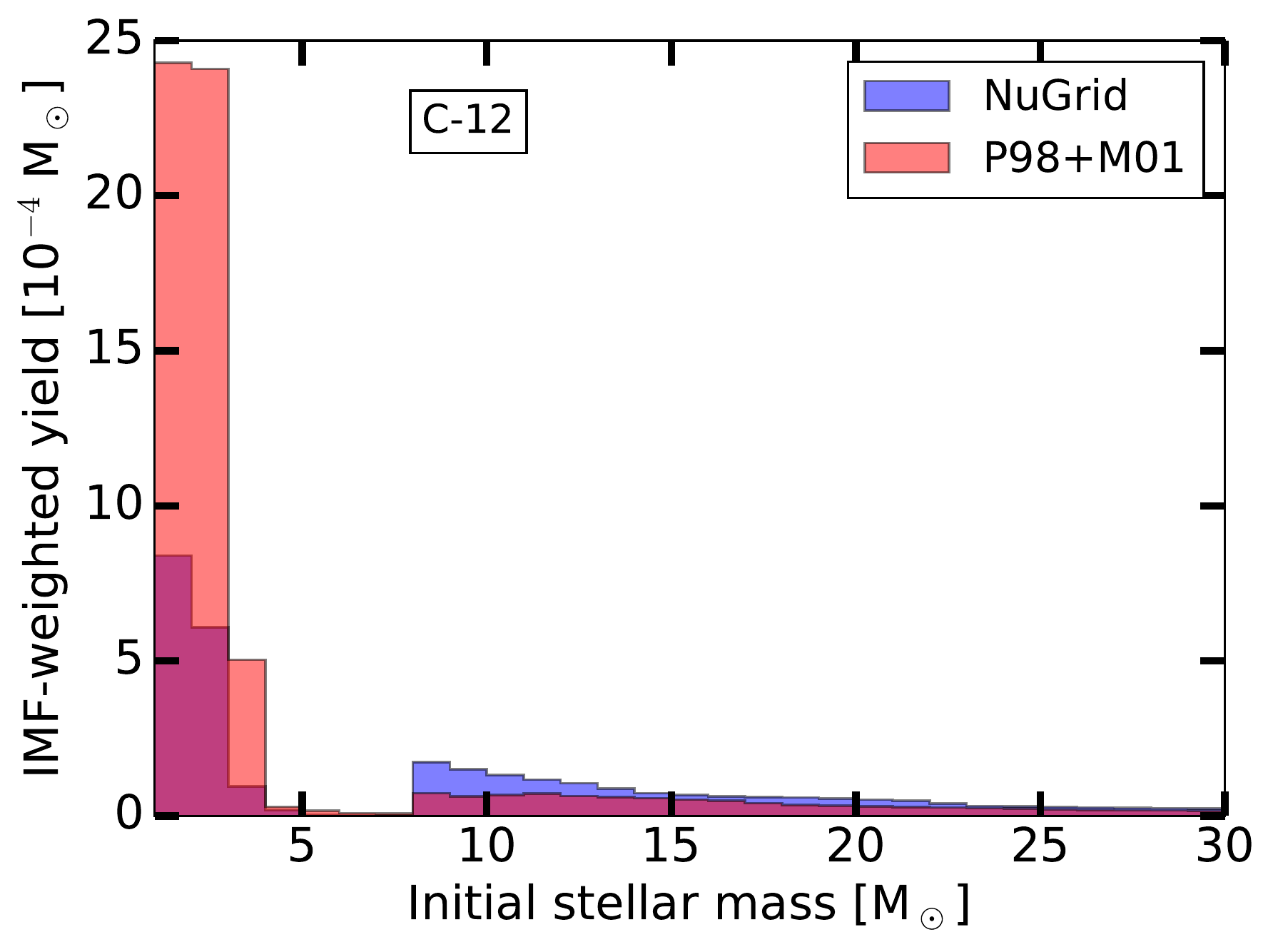}\hspace*{0.1cm}
\includegraphics[width=0.49\textwidth]{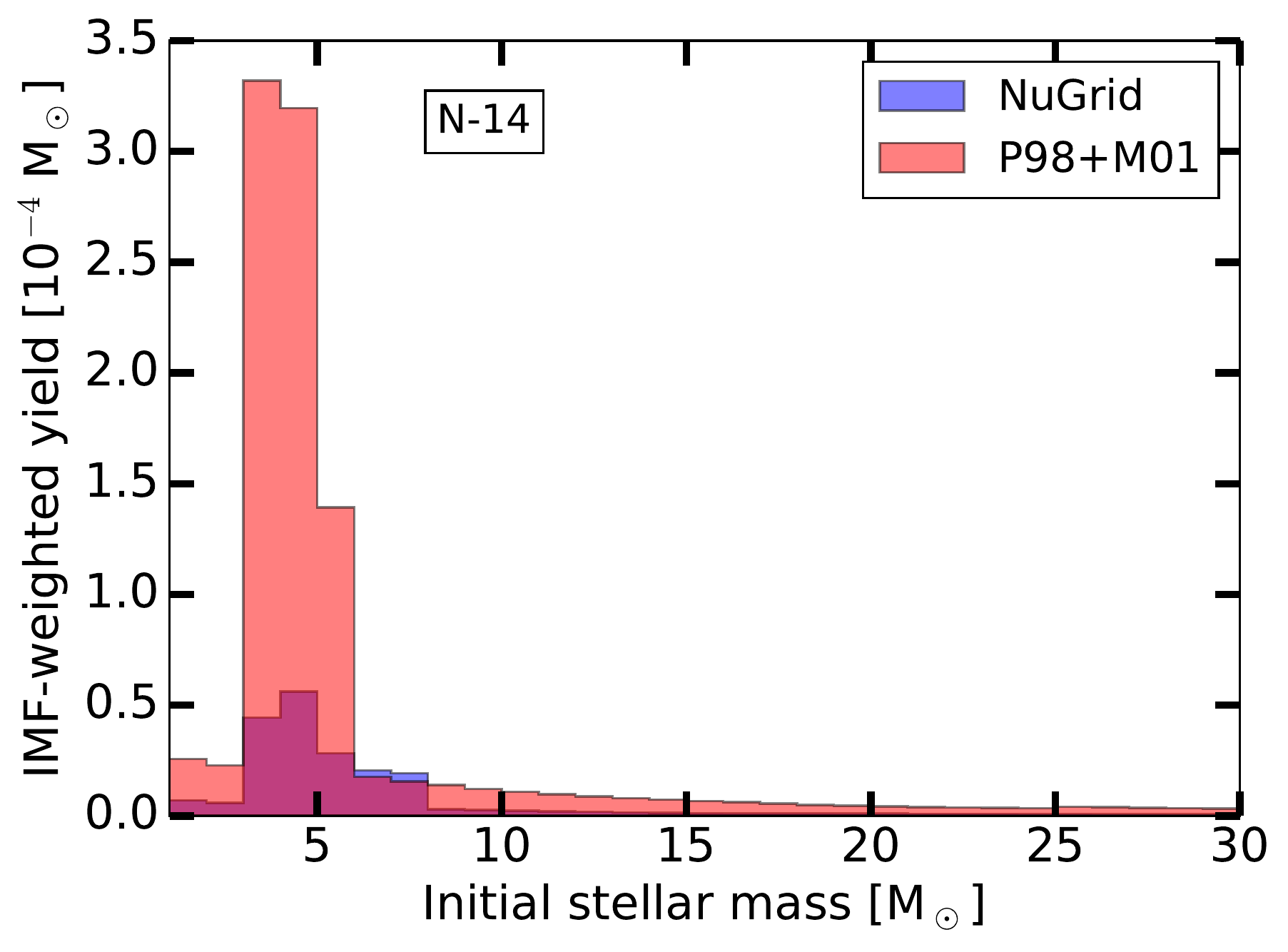}
\caption{IMF-weighted yields of C represented by \isotope[12]{C} and N represented by \isotope[14]{N} versus initial mass based on NuGrid
yields with $Z=0.001$ and M01P98 yields with $Z=0.004$, for an SSP of $1\msun$.
}
\label{fig:ssplowz_c12n14}
\end{figure*}

\paragraph{C, N} 
The total SSP ejecta of C from AGB stars is 71\% lower with NuGrid
yields compared to yields of M01P98 (Figure~\ref{fig:ssplowz}).  This
is due to the low-mass AGB models of P98 which eject larger amounts of
C compared to NuGrid models (Figure~\ref{fig:ssplowz_c12n14}).  M01
use synthetic models which do not model the TDUP self-consistently
contrary to the NuGrid models.  In the synthetic models the dredge up
of C into the envelope results from the calibration of the dredge-up
strength and minimum core mass of TDUP occurrence against the observed
carbon star luminosity distribution.  The total SSP ejecta of C of
massive stars is with NuGrid yields 37\% higher than with M01P98
yields.

The total AGB ejecta of N of the SSP based on NuGrid yields is 80\%
lower than the N ejecta based on M01P98 yields.  We find large
discrepancies between N yields from NuGrid and yields of M01 for
massive AGB models (Figure~\ref{fig:ssplowz_c12n14}).  The amounts of
N produced in these stellar models during HBB depend on the length of
the TP-AGB phase which is based on free parameters of the synthetic
models of M01. The mass loss model adopted in the NuGrid
  simulations does also include an uncertain efficiency parameter. N
  yields in the NuGrid models are a result of convective boundary
  mixing assumptions and the modeling of the third dredge-up
  \citep[see][for details]{ritter:17}. The SSP ejecta of N of massive
stars with NuGrid yields is for most of the evolution 80\% lower than
with yields by M01P98.

\paragraph{O, Si} 
The stellar yields of O from AGB models differ more between the yield
sets at lower metallicity than at solar metallicity.  The difference
is the largest for the most massive AGB models. This translates in a
large difference of AGB ejecta of the SSPs early on in the evolution
(Figure~\ref{fig:ssplowz}).  With NuGrid yields the total amount of O
ejected by AGB stars is 12\% below the amount ejected with M01P98
yields.  The evolution of SSP ejecta of O from massive stars is in
slightly better agreement between the yield sets at low metallicity
than at solar metallicity.  The total SSP ejecta of O of massive stars
with NuGrid yields is 39\% lower than with M01P98 yields.

There is little SSP ejecta of Si from AGB stars.  The
  difference between M01P98 and NuGrid yields in the Si ejecta is due
  to the different initial abundances of the stellar models.  NuGrid
adopts an $\alpha$-enhanced initial abundance for models of $Z=0.001$.
When we apply the same initial abundance of Si AGB ejecta of both sets
are in good agreement.  The SSP ejecta of Si from massive stars are
similar for the yield sets as at solar metallicity.  We find a 28\%
lower total amount of Si ejected by massive stars of NuGrid compared
to M01P98.

\paragraph{Fe} 
As for Si, the Fe ejecta from AGB stars based on NuGrid yields
  are lower than with M01P98 yields due to the adopted initial
  abundances.  SSP ejecta of Fe of massive stars based on NuGrid
yields starts below and then increases above the ejecta based on
M01P98 yields.  The strong fallback in NuGrid models of high initial
mass limits the Fe ejecta at early times in the SSP evolution. The
total amount of Fe ejected is 23\% larger with NuGrid yields than with
M01P98 yields.

\subsection{Impact of Core-collapse Mass-cut Prescriptions} \label{s.science5}

\begin{figure*}[!htbp]
\centering
\includegraphics[width=1.\textwidth]{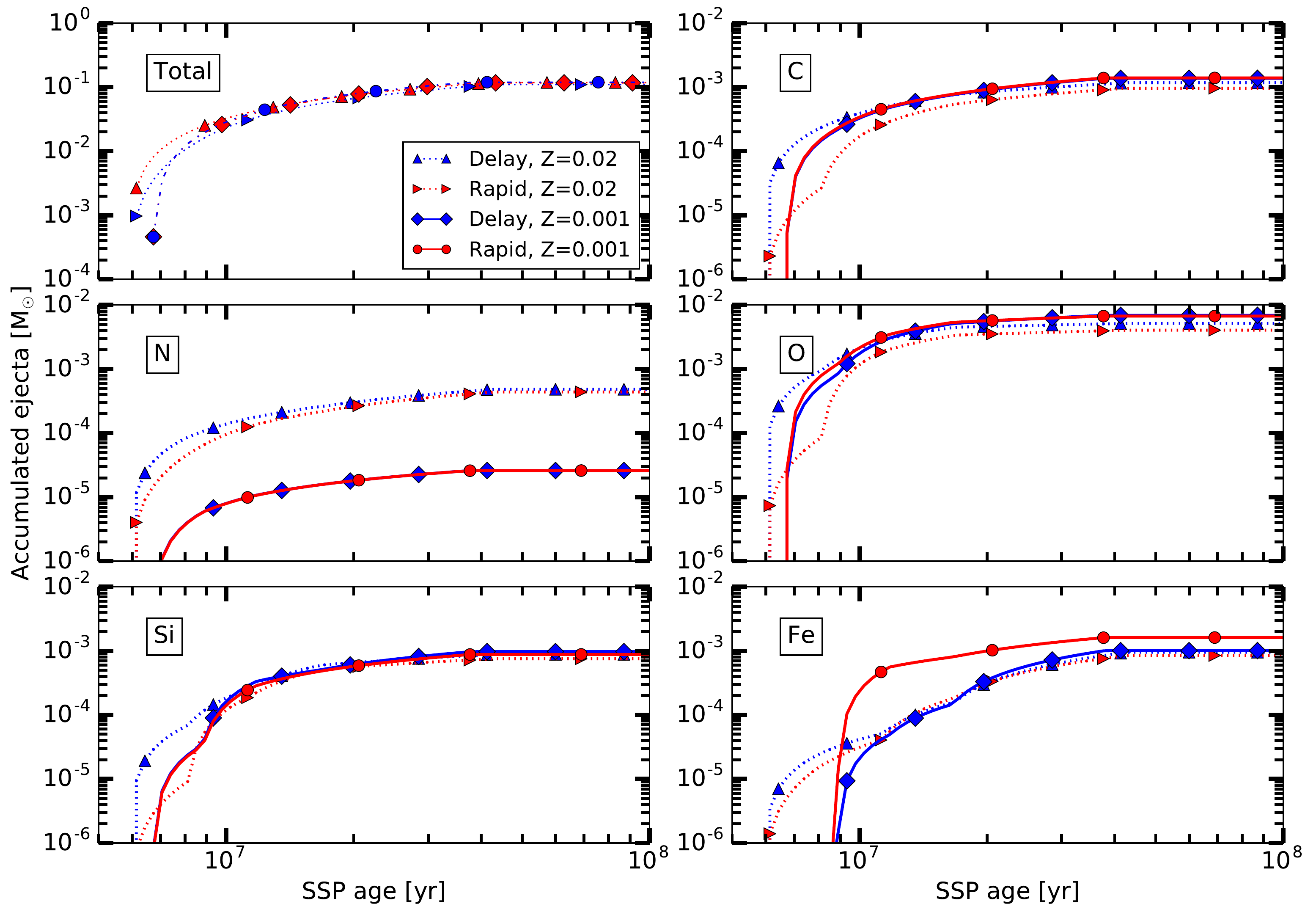}
\caption{Accumulated ejecta of massive stars of NuGrid yields at $Z=0.02$ and $Z=0.001$ computed with
the delayed and rapid CCSN fallback prescription, for an SSP of $1\msun$.
}
\label{fig:ssp_fallback}
\end{figure*}

One of the major uncertainties in calculating yields for SSPs is the
treatment of fallback after supernova explosions. NuGrid provides
two explosion prescriptions based on the
convective-enhanced neutrino-driven engine.
\citep[e.g.][F12]{fryer:07,fryer:12}.  The rapid and delayed
  explosion assumptions refer to the case where a short and long time
  are needed for the convective engine to revive the
  shock. Accordingly, delayed explosions have generally more time for
  fallback and larger remnants, and rapid explosions have smaller remnants.
  The SSP ejecta of C, O and Si differ more between the fallback
prescriptions at $Z=0.02$ than at $Z=0.001$ due to the BH formation
after the rapid explosion of the stellar model with $\mzams=25\msun$
at $Z=0.001$ \citep[see][for details]{ritter:17} which prevents the
ejection of the Si, O and C shell (Figure~\ref{fig:ssp_fallback}).

For the lower metallicity case the Fe ejection is larger for the rapid
explosion as more Fe falls back in the delayed explosion case.  Within
the first $1.3\times10^7\, \mathrm{yr}$ ten times more Fe is ejected
with the rapid explosion compared to the delayed explosion. The
expected range of explosion delay times (F12) could lead to large
variations in Fe enrichment in the early universe.

The rapid explosion models match the observed gap between NS and BH
remnants better than the delayed models even though the gap might be
sparsely populated (F12 and references therein).  Delayed explosions
produce more fallback BHs, in particular at low mass and yield a
larger fraction of low-mass BHs formed with a SN explosion.  The
latter can explain the observed BH systems which indicate a natal kick
(F12).  Fallback is also relevant to produce the weak supernova
\citep{valenti:09} which are believed to be observed (F12).  We
recommend yields with the delayed explosion prescription in chemical
evolution models. However, both cases may be necessary to explain all
SN observations (F12).  Therefore NuGrid offers yield tables that
contain a half and half mixture of rapid and delayed options.

% !TEX root = ./paper.tex
\section{Summary and Conclusions}\label{s.conclusions}

%\noindent 
\code{SYGMA} provides the chemical ejecta and stellar feedback of
simple stellar populations for application to galactic chemical
evolution, hydrodynamical simulations and semi-analytical models of
galaxies.  A variety of SNIa delay-time distributions, IMF and yield
options are included. \code{SYGMA} includes various non-standard
nucleosynthesis sources, such as NS mergers and CCSN neutrino-driven
winds, and can track the corresponding r-process enrichment.  Along
with the built-in plotting and analysis tools and the online
web-accessible version (Appendix\ \ref{s.appendixA}) \code{SYGMA} can
be used to trace cumulative features of GCE models to individual
nuclear astrophysics properties of individual stellar
models. \code{SYGMA} is part of the NuGrid Pyhon Chemical Evolution
Environment \code{NuPyCEE} and consitutes a key building block in
the JINA-NuGrid chemical evolution pipeline
\citep[e.g.][]{cote:17c,cote:17d} that integrates fundamental stellar
and nuclear physics investigations with galactic modeling
applications.

 The primary focus of \code{SYGMA} is on investigating and applying
 the present and future NuGrid yields that are calculated using the
 same simulation tools and -- as much as possible and appropriate --
 physics assumptions for low-mass and massive star models that are
 based on stellar evolution calculations and comprehensive
 nucleosynthesis simulations.  Large, unquantifiable uncertainties that
 originate from the application of different AGB and massive star
 yield sets are avoided. Stellar feedback from CCSNe is consistent
 with the underlying explosion models of the NuGrid data set.
 Stellar luminosities in energy bands are calculated from
 time-dependent synthetic spectra of NuGrid stellar models.  \code{SYGMA} tracks an arbitrary number of
 elements and isotopes up to Bi as provided by the NuGrid
yields.
 
NuGrid yield tables are part of the
   larger \code{NuPyCEE} yield library which also includes a number of
   other yield sets from the literature, as well as yields from more
   exotic sources, such as MHD jets (\citealt{nishimura:15}),
   hypernovae (\citealt{kobayashi:06}), and NS mergers
   (\citealt{rosswog:14}). Additional tables can be added on request.

A comparison of SSPs using NuGrid yields and the combined yields of
P98 and M01 expose differences for C and N ejecta due to massive and
super-AGB star yields and are the largest at low metallicity with a
factor of 3.5 and 4.8.  Different CCSN fallback treatments result in
differences in C, O and Si of up to a factor of ten in certain cases,
such as the first $10^7 \mathrm{yr}$ of an SSP at $Z=0.001$ are
possible. The largest difference of the total ejecta between the
fallback prescriptions is for Fe with a factor 1.6. A brief code
comparison of \code{SYGMA} with the W09 SSP code demonstrates the code
design and implementation impact (\sect{s.appendixveri}).

%results
The functionality of the module was verified through a comparison with W09 in which we apply their yields.
The final accumulated ejecta are well in agreement in both works and the largest differences in the fraction of ejecta
of N and Fe from massive stars is 10\%.

% and the basic building block
%of the galactic chemical evolution code OMEGA \citep{}.

\acknowledgments
%\noindent 
We are thankful to the anonymous referee for providing valuable feedback on this work.
We are grateful to Else Starkenburg and Kim Venn for their valuable discussion about the concept of the \code{SYGMA} module.
%valuable discussions about the implementation. 
We are thankful to Luke Siemens and Jericho O'Connell for building the IPython widgets for \code{SYGMA}.
We acknowledge Adam Paul for his contributions to the NS merger implementation.
NuGrid acknowledges significant support from NSF grant PHY-1430152 (JINA Center for the Evolution of the Elements). BC acknowledge supports from the FRQNT (Quebec, Canada) postdoctoral fellowship program and the ERC Consolidator Grant (Hungary) funding scheme (project RADIOSTAR, G.A. n.\,724560).
We acknowledge the Canadian Advanced Network for Astronomical Research cloud service which hosts the \code{SYGMA} web interface. %github, ipython notebooks??
CR received funding from the European Research Council under the European Union's Seventh Framework Programme (FP/2007-2013)/ERC Grant Agreement n. 306901.

\software{\texttt{SYGMA} (this work), \texttt{NuPyCEE}
  \citep{2016ascl.soft10015R}, \texttt{Cyberhubs}
  \citep{2018ApJS..236....2H}, \texttt{NumPy} \citep{2011arXiv1102.1523V},
  \texttt{matplotlib} (\url{https://matplotlib.org}).}

\bibliographystyle{yahapj}
\bibliography{apj-jour,astro}

% for appendices, if you have them
\appendix

\section{Code verification}
\label{s.appendixveri}
We compare \code{SYGMA} results calculated with the same yields as W09
(M01P98) with the W09 results.  This serves two goals. The first is
simply code verification. The second is to provide some estimate on the
kinds of uncertainties that are introduced by small code design and
implementation differences, which are in addition to the uncertainties
in the yield input data. We choose for this comparison the widely used
W09 work, but would expect to find similar outcomes when comparing
with other SSP codes. As this appendix shows, the differences are small
but not entirely insiginficant. Our response to these differences is
to make our code public so that all code design and implementation
details can be scrutinized, and changed if deemed appropriate.

We apply the same massive star model factors of 0.5, 2 and 0.5 as in
W09, initial abundances from Table 1 of W09, and -- as much as
possible -- the same chemical evolution parameters
(\tab{tab:ce_parameter}). The initial mass range for which yields were
ejected is not given W09 and we choose the range from
$\mzams=0.8\msun$ to $\mzams=100\msun$ to match best Fig.\ 2 in W09.
\begin{figure*}[!htbp]
\centering
\includegraphics[width=1.\textwidth]{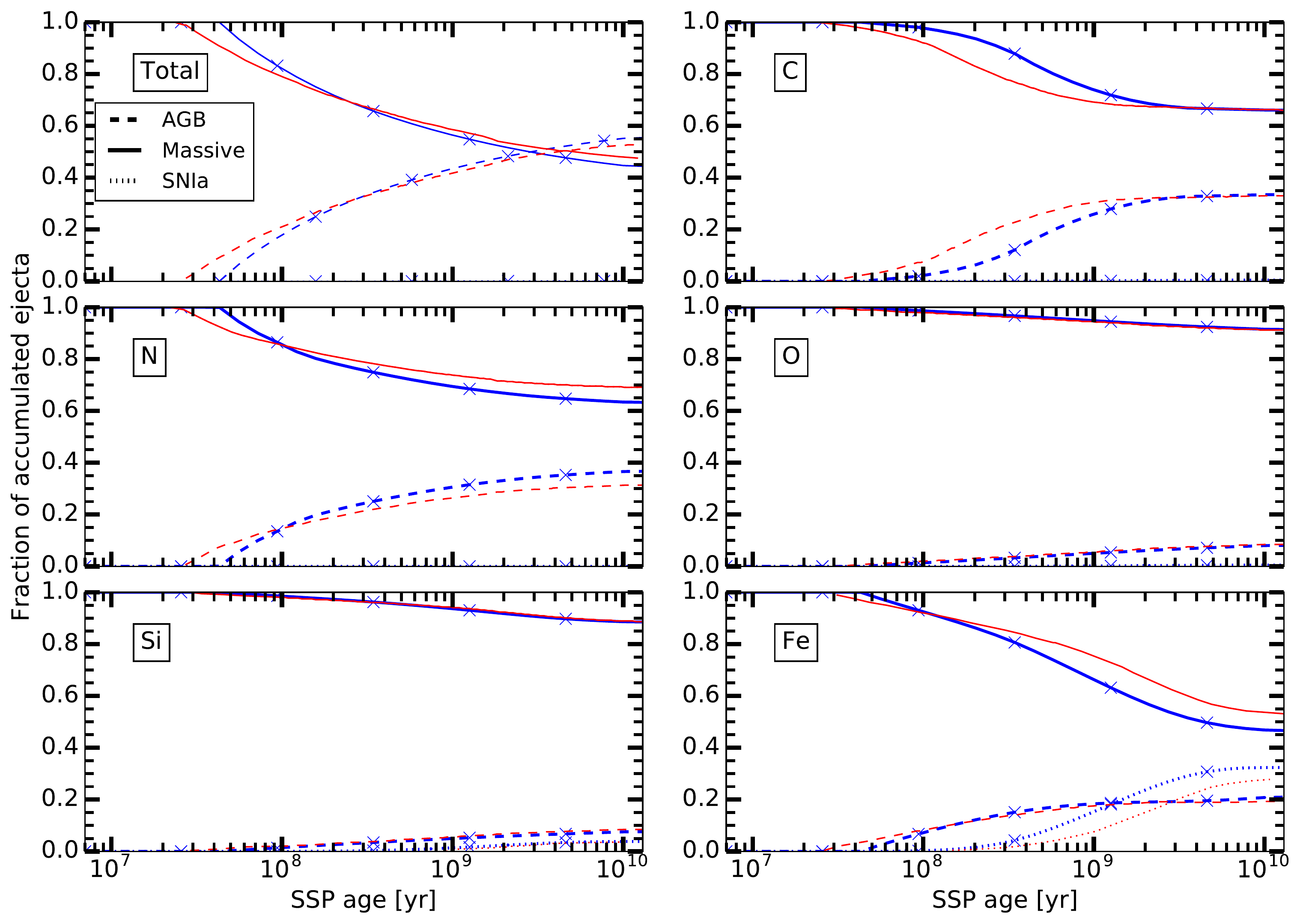}
\caption{Fraction of total mass ejected from AGB, massive star and SNe
  Ia for a SSP at solar metallicity with yield input from M01P98
  (blue, crosses) compared to results extracted from Fig.\ 2 in W09
  (red).  Results are based on the same yield input but different SSP
  codes.  }
\label{fig:wiersma_fig2}
\end{figure*}

Another important parameter is the transition initial mass $M_{\rm
  mass}$ that delineates white dwarf and supernova outcomes. It is not
given in W09, but must be between $\mzams=7$ and $9\msun$ in the
M01P98 set. The actual value of $M_{\rm mass}$ is still a matter of
some debate \citep{poelarends:08,doherty:15,doherty:17,jones:16}. For
this section we have adopted a value (\tab{tab:ce_parameter}) that
agrees best with the results shown in Fig.\ 2 of W09.

\begin{figure}[!htbp]
\begin{center}
\includegraphics[width=0.95\textwidth]{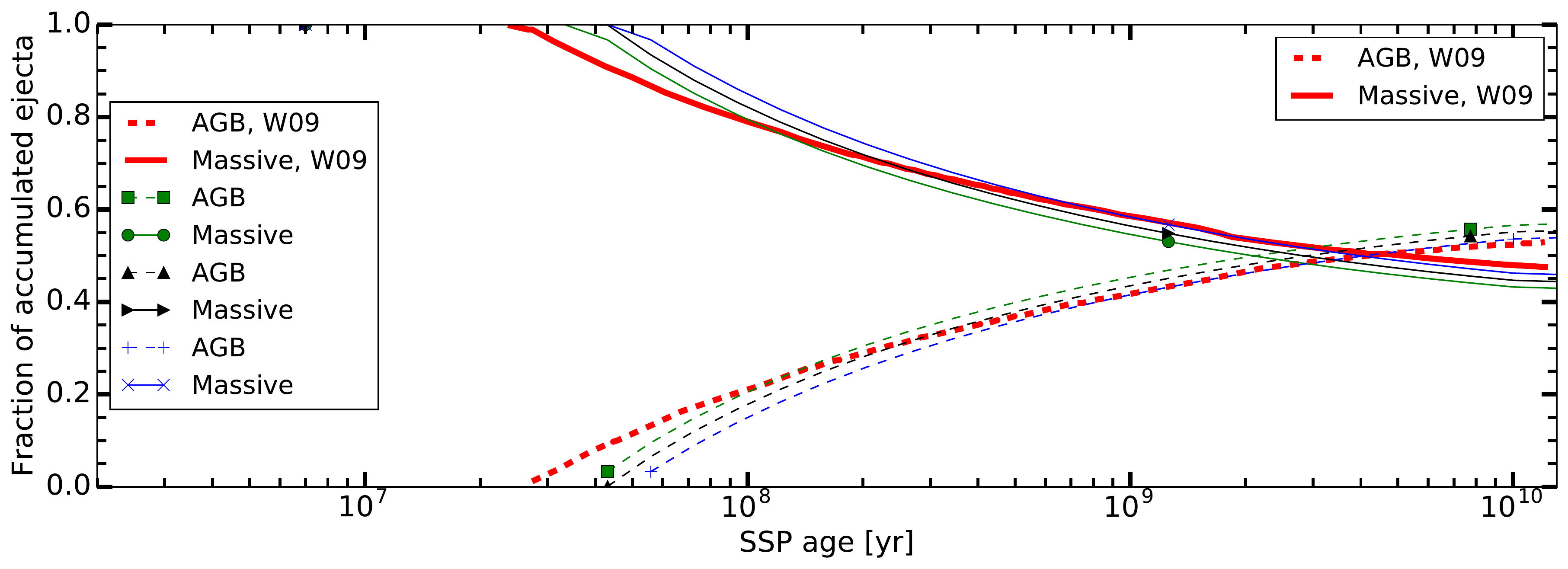}
\caption{Evolution of fraction of total ejecta for transition masses of $\mzams=7.5\msun$, $\mzams=8\msun$ and $\mzams=8.5\msun$. 
}
\label{fig:ce_comparisons}
\end{center}
\end{figure}
Overall the \code{SYGMA} and W09 SSP models agree well, but differences can
be seen as low- and intermediate mass stars start to contributed --
especially for C (up to $10\%$ differnce) and Fe
(Figure~\ref{fig:wiersma_fig2}).  The choice of the transition mass
$M_{\rm mass}$ determines the appearance of the AGB star ejecta and
the drop in total massive star ejecta
(Figure~\ref{fig:ce_comparisons}).
The N yields increase smoothly with initial mass which leads to a smooth increase of the SSP ejecta similar to W09.
The differences in the C and N evolution could be due to different yield interpolation methods used in the initial mass transition region from AGB to massive stars.

\section{Online availability}
\label{s.appendixA}
The \code{SYGMA} web interface allows to simulate, analyse, and extract
SSP ejecta which includes all stable elements and many isotopes up to Bi.
We introduce the yield sets and parameters which are available within the web interface.
Yields for AGB stars and massive stars can be selected from the NuGrid sets NuGrid$_{\rm d/r/m}$ (Table~\ref{tab:yieldcompilations})
SNIa yields are from \cite{thielemann:03} and Pop III yields are from \cite{heger:10}.
The available metallicities are $Z=0.02$, $0.01$, $0.006$, $0.001$ and $0.0001$, $0$.
Yields are applied in the initial mass range from $1\msun$ to $30\msun$.
Chemical evolution parameters such as IMF and SNIa DTD can be set.

SSP ejecta can be extracted in the form of tables which contain for each time step the fraction of elements and isotopes of choice.
As an example parts of a table which contains the normalized mass of elements ejected over $10^{10}\, \mathrm{yr}$ by a SSP of $1\msun$ at $Z=0.02$ 
is presented in Table~\ref{tab:abu_table}.

%%%%%%%%%
\begin{deluxetable}{lccccccc}
\tablewidth{0pc}
\tablecaption{Sample of a table including time evolution of the ejected elements extracted with the \code{SYGMA} web interface, for a SSP of $1\msun$ at $Z=0.02$. \label{tab:abu_table}}
\tablehead{ \colhead{Age [yr]}  & \colhead{C}         &\colhead{N}         &\colhead{O}         &\colhead{Fe}        &\colhead{Sr}        &\colhead{Ba}   &     \colhead{M$_\mathrm{tot}$ [\msun]} }
\startdata
1.000E+07 &3.921E-04 &1.391E-04 &2.118E-03 &4.179E-05 &1.268E-08 &7.086E-10 &3.273E-02 \\
\multicolumn{8}{c}{......} \\
1.000E+08 &1.208E-03 &8.412E-04 &5.497E-03 &1.025E-03 &3.713E-08 &3.192E-09 &1.642E-01 \\
\multicolumn{8}{c}{......} \\
1.000E+09 &2.629E-03 &1.212E-03 &7.025E-03 &1.301E-03 &5.625E-08 &7.224E-09 &2.793E-01 \\
\multicolumn{8}{c}{......} \\
1.000E+10 &3.257E-03 &1.408E-03 &8.052E-03 &1.655E-03 &6.211E-08 &9.051E-09 &3.691E-01\\
\enddata
\end{deluxetable}
%%%%%%%%%

The \code{SYGMA} code and the yield library can be accessed via
\href{http://nugrid.github.io/NuPyCEE/}{\url{http://nugrid.github.io/NuPyCEE}}.
We provide an online documentation based on
SPHINX\footnote{http://www.sphinx-doc.org}, guides and teaching
material in form of Jupyter notebooks.  \code{SYGMA} web interface is
accessible through the NuPyCEE web page and hosted on NuGrid's Web
Exploration of NuGrid Datasets: Interactive (WENDI) platform at
\url{http://wendi.nugridstars.org}. WENDI is a \code{Cyberhubs}
service \citep{2018ApJS..236....2H}. Access to the figures of this work are
provided through WENDI.  NuGrid's stellar and nucleosynthesis data
sets are available at
\url{http://nugridstars.org/data-and-software/yields/set-1} and can be
analyzed with WENDI.

\end{document}